\documentclass[twocolumn,reprint, english,aps,prb,superscriptaddress]{revtex4-1}
\usepackage{amsmath}
\usepackage{amssymb}
\usepackage{esint}
\usepackage{comment}

\usepackage{graphicx}
\makeatletter
\@ifundefined{textcolor}{}
{%
 \definecolor{BLACK}{gray}{0}
 \definecolor{WHITE}{gray}{1}
 \definecolor{RED}{rgb}{1,0,0}
 \definecolor{GREEN}{rgb}{0,1,0}
 \definecolor{BLUE}{rgb}{0,0,1}
 \definecolor{CYAN}{cmyk}{1,0,0,0}
 \definecolor{MAGENTA}{cmyk}{0,1,0,0}
 \definecolor{YELLOW}{cmyk}{0,0,1,0}
}

\usepackage{babel}

\makeatother

\usepackage{color}
\usepackage{graphicx}
\usepackage[colorlinks=true,citecolor=blue,linkcolor=blue]{hyperref}

\begin{document}

\title{Universal post-quench coarsening and quantum aging at a quantum critical point }

\author{Pia Gagel}
\affiliation{Institute for Theory of Condensed Matter, Karlsruhe Institute of
Technology (KIT), 76131 Karlsruhe, Germany}

\author{Peter P. Orth}
\affiliation{Institute for Theory of Condensed Matter, Karlsruhe Institute of Technology (KIT), 76131 Karlsruhe, Germany}
\affiliation{School of Physics and Astronomy, University of Minnesota, Minneapolis, Minnesota 55455, USA}

\author{J\"org Schmalian}
\affiliation{Institute for Theory of Condensed Matter, Karlsruhe Institute of
Technology (KIT), 76131 Karlsruhe, Germany}
\affiliation{Institute for Solid State Physics, Karlsruhe Institute of Technology (KIT), 76131 Karlsruhe, Germany}

\date{\today}
\begin{abstract}
The non-equilibrium dynamics of a system that is located in the vicinity
of a quantum critical point is affected by the critical slowing down
of order-parameter correlations with the potential for novel out-of-equilibrium universality. After a quantum quench, i.e. a sudden change of a parameter in the Hamiltonian such a system is expected to almost instantly fall out of equilibrium and undergo aging dynamics, i.e. dynamics
that depends on the time passed since the quench. Investigating the
quantum dynamics of a $N$-component $\varphi^{4}$-model coupled
to an external bath, we determine this universal aging and demonstrate
that the system undergoes a coarsening, governed by a critical exponent
that is unrelated to the equilibrium exponents of the system. We analyze
this behavior in the large-$N$ limit, which is complementary to our
earlier renormalization group analysis, allowing in particular the
direct investigation of the order-parameter dynamics in the symmetry
broken phase and at the upper critical dimension. By connecting the long time limit of fluctuations and response, we introduce a distribution function that shows that the system remains non-thermal and exhibits quantum coherence even on long timescales. 
\end{abstract}
\maketitle

\section{Introduction}
\label{sec:introduction}
Non-equilibrium behavior of interacting many-body systems is becoming a field of increasing importance in different areas of physics. It is largely driven by two main experimental advances: (i) the ability to bring a system into an out-of-equilibrium state in a controlled and reproducable manner and (ii) the ability to observe and follow the subsequent dynamics in real-time, i.e. on microscopic timescales. Different tools have been developed for different experimental systems. Important examples are various cold-atom setups~\cite{bloch:885,RevModPhys.83.863,morsch:179,kinoshita_nature_2006, Gring14092012, LangenGeigerSchmiedmayer-AnnRevCondMatt-2015}, ultra-fast pump-probe spectroscopy of quantum materials~\cite{Orenstein-PhysToday-2012, Fausti14012011, Smallwood01062012, HuCavalleriKeimer-NatPhys-2014}, and heavy-ion collisions that explore the dynamics of the quark-gluon plasma~\cite{Arsene20051}. These experiments clearly reveal that observations far away from equilibrium can yield new insights. 

Non-equilibrium dynamics is often characterized by a lack of time-translational invariance. As a result, the fluctuation-dissipation theorem does typically not hold and memory effects occur, i.e. the dynamics depends on the initial state of the system. Major questions in non-equilibrium are (i) what are the effects of interactions; do non-linearities, aging and memory effects occur and how are they characterized~\cite{Janssen-ZPhysB-1989, 0305-4470-38-18-R01, BouchaudCugliandolo-AgingInBook-1997}, (ii) what are the properties of transient metastable ``pre-thermal'' states~\cite{PhysRevLett.93.142002, 1367-2630-13-7-073018, PhysRevB.84.054304, PhysRevLett.100.175702, PhysRevB.75.144418, PhysRevLett.103.056403, 1367-2630-12-5-055008, PhysRevB.87.205109, PhysRevLett.106.050405, kollath:180601,PhysRevLett.98.210405,PhysRevLett.111.197203,PhysRevLett.110.136404}, and (iii) how does the system eventually reach a steady-state and is it given by a (generalized) thermal distribution~\cite{PhysRevA.43.2046,PhysRevE.50.888,Rigol-Thermalization-Nature-2008,RevModPhys.83.863}. Aging describes the phenomenon that correlation and response functions $G^K(t,t')$  and $G^R(t,t')$ depend on both time arguments $t,t'$ and not just on their time difference $t-t'$ as is the equilibrium case, i.e. they depend the age of the system. These effects are well-known to occur in structural glasses, spin-glasses and disordered systems~\cite{FischerHertz-SpinGlasses-1991, LubchenkoWolynes-StructuralGlassesReview-JChemPhys-2004, PhysRevLett.96.217203}. As these effects are related to spatial and temporal fluctuations it is often required to perform an analysis beyond the mean-field approximation. 

It is in general difficult to make quantitative predictions for the non-equilibrium dynamics of interacting quantum many-body systems beyond the mean-field approximation~\cite{RevModPhys.86.779}- except in cases where exact solutions are available~\cite{PhysRevLett.110.135704,0295-5075-85-2-20004, PhysRevB.88.104511, 1742-5468-2012-07-P07016, 1742-5468-2012-07-P07022, PhysRevLett.106.227203, PhysRevLett.96.097005, PhysRevLett.96.230404, PhysRevLett.80.4370,PhysRevB.88.205131, DzeroKhodasLevchenko-QuenchInSc-arxiv-2015} or when controlled numerical approaches are established such as for impurity models~\cite{PhysRevLett.95.196801, PhysRevLett.102.196601, PhysRevB.82.144423, PhysRevLett.104.106801, PhysRevB.77.195316,PhysRevB.78.235110,PhysRevB.87.014305,PhysRevB.85.085113} and for one-dimensional systems~\cite{Giamarchi-QuantumPhysIn1D,PhysRevLett.69.2863, RevModPhys.77.259,  PhysRevB.70.121302, PhysRevLett.93.076401, PhysRevLett.100.100601, PhysRevLett.113.010601, PhysRevB.87.205109,PhysRevLett.109.260601,2014arXiv1404.3740B}. Here, we exploit the presence of a quantum critical point to analytically solve for the universal dynamics of an interacting quantum $\varphi^4$-model, which is an effective description of a number of experimental systems~\cite{sachdev_qpt_book}. We obtain solutions both for short as well as for long times. 

Universality close to (quantum) critical points is well established in equilibrium and follows from a divergence of the correlation length $\xi$ and time $\xi_\tau$ near criticality~\cite{sachdev_qpt_book,RevModPhys.69.315}. Observables such as the order parameter or correlation functions can be expressed in terms of universal scaling functions with dimensionless arguments. As a function of, for example, the distance to the critical point they follow power-laws with universal critical exponents. In non-equilibrium, the correlation length is itself a function of time. For example, if a parameter of a system is changed sufficiently slowly, the state and the correlation length of the system can adiabatically follow this change. Close to a critical point, however, adiabaticity demands that the parameter change must occur infinitesimally slowly due to the divergence of the correlation time. At any finite sweep rate, the system eventually falls out of equilibrium. In the Kibble-Zurek description of such a parameter sweep through a critical point, the correlation length is assumed to remain constant at this freeze-out length scale for the remainder of the sweep~\cite{Kibble-JPhysA-1976,Zurek-Nature-1985}. It then follows that the number of (topological) excitations depends on the rate via a universal scaling law that solely contains equilibrium critical exponents~\cite{PhysRevLett.95.035701,0295-5075-84-6-67008, PhysRevLett.109.015701, PhysRevB.86.064304,PhysRevB.84.224303}. Similarly, it follows that the long time approach to equilibrium of a system that is suddenly quenched close to a (quantum) critical point is governed by equilibrium exponents~\cite{PhysRevB.81.012303}. We confirm this behavior in our study as well. Kibble-Zurek scaling has been experimentally observed, in particular in the regime of slow sweeps~\cite{RevModPhys.83.863,PhysRevLett.83.5210, WeilerAnderson-KZMInBEC-Nature-2008, Navon09012015}. In the opposite regime of fast parameter quenches also different scaling behavior has been reported~\cite{PhysRevX.2.041022}. 

Recently, however, it was shown for the KZ regime of a gradual sweep, that the dynamics of a system does not completely freeze out once the system falls out of equilibrium~\cite{PhysRevE.81.050101}. Instead, the correlation length $\xi(t)$ remains a function of time. If the system is located close to a critical point, where $\xi$ diverges in equilibrium, the correlation length in fact diverges in a light-cone like fashion according to a power law 
\begin{align}
  \label{eq:3}
  \xi(t) \propto t^{1/z_d}
\end{align}
with a characteristic dynamic coarsening exponent $z_d$~\cite{PhysRevE.81.050101}. While the coarsening exponent is in general different from the dynamic critical exponent $z$ that characterizes the quantum dynamics in equilibrium, it turns out that the two are equal in our approach. This scaling behavior~\eqref{eq:3} is also one of the results of our calculations for the case of a quantum quench. A crucial implication of it is that correlation and response functions $G^K, G^R$ obey a generalized scaling form, which incorporates the time dependence of $\xi(t)$. Using this scaling form, we are able to find an analytic solution of the universal dynamics of correlation and response functions, both at short and at long times. 

In this article we study the post-quench non-equilibrium dynamics of a dissipative $N$-component $\varphi^4$-model. Dissipation is introduced via coupling to an external bosonic bath with ohmic, sub-ohmic or super-ohmic spectrum. In equilibrium, the model exhibits a quantum critical point (QCP) that separates a disordered (symmetry unbroken) from an ordered (symmetry-broken) phase. The equilibrium properties of this model are discussed, for example, in Refs.~\onlinecite{PhysRevB.14.1165, PhysRevB.48.7183, sachdev_qpt_book}. We consider a system that is initially prepared in an equilibrium state away from the quantum critical point, where the correlation length $\xi$ is finite. We discuss both the situation of an initial state in the symmetry-broken and in the symmetry-unbroken phase. Then, at time $t=0$ a parameter in the system is rapidly changed, i.e. quenched, to a value that corresponds to the quantum critical point in equilibrium. Experimentally, this parameter can, for example, be pressure, strain, magnetic field or interaction strength.

We are interested in the post-quench dynamics of the correlation and response functions as well as the time evolution of the order parameter itself. We obtain analytical expressions for the universal part of the time dependence. Due to the rapid quench the correlation length first collapses to a non-universal value of the order of a microscopic length scale in the system. Since the system is located at the quantum critical point, however, it then recovers by critical coarsening in the light-cone fashion $\xi(t) \propto t^{1/z_d}$. An important implication of this power-law recovery of $\xi$ is a universal prethermalized regime where the dynamics of the correlation function and the order parameter $\phi(t)$ is described by the universal critical exponent 
\begin{align}
  \label{eq:4}
  \phi(t) \propto t^\theta\,.
\end{align}
For $\theta > 0$ follows that the order parameter increases after an initial collapse due to the fast spreading of correlations in the system, while for $\theta<0$ the order decreases. The duration of the prethermalization regime ix determined by the quench amplitude and increases for smaller quench amplitudes, i.e. if the initial state of the system was close to criticality. We note that while universality is often associated with large length and time scales, it here arises from the critical light-cone recovery of the correlation length over time, similar to the coarsening in heterogeneous systems~\cite{PhysRevLett.49.1545}.  After the prethermalized regime at large times, we show that the system eventually relaxes to equilibrium quasi-adiabatically with a power-law that contains equilibrium critical exponents. Still, the prethermalization exponent $\theta$ enters the long-time expressions as a universal pre-factor in the relaxation amplitudes of the order parameter and the correlation functions. We note that the dominant slow dynamics is determined by the interactions that are present within the model and the coupling to the bath mainly ensures equilibration in the long time limit. At long limes we can define a time-dependent distribution function $n(t, \omega)$ using the relation between response and correlation functions known as the fluctuation-dissipation theorem in equilibrium. We find that the deviation $\delta n $ from the equilibrium Bose-Einstein distribution is given by
\begin{equation}
 \delta n(t,\omega)=-\coth\left(\frac{\omega}{2T} \right) \frac{\gamma \theta\Gamma(2/z)}{\sin(\pi/z) t^{2/z}} \text{Re} \, G^R_{\text{eq}}(q,\omega),
\end{equation}
where $\gamma$ is the coupling to the bath, $T$ the bath-temperature, $G^R_{\text{eq}}$ denotes the equilibrium retarded Green's funtion at the QCP and $t$ is the time after the quench. As we show, this result implies that the distribution function is non-thermal even in the long time regime: $\delta n$ decays only algebraically at large frequencies, while a thermal distribution decays exponentially. The approach to equilibrium over time is slow and described by a power-law $\propto t^{-2/z}$. Finally, the fact that $\delta n$ changes sign and can become negative for $z < 2$ implies that the density matrix of the system is non-diagonal in the energy basis and quantum coherence is present.  

The quench protocol to the critical point that we consider is inspired by the pioneering work on classical critical points by Janssen, Schaub and Schmittmann in Ref.~\onlinecite{Janssen-ZPhysB-1989} who analyzed the post-quench dynamics of a classical $\varphi^4$ theory in contact with an ohmic bath. The analysis of the classical case was extended to colored noise in Ref.~\onlinecite{1742-5468-2012-01-P01014}. Previously, we have reported results on the dissipative quantum $\varphi^4$-model using a renormalization-group (RG) approach in non-equilibrium and calculated the new exponent $\theta$ in an expansion in $\epsilon = 4 - d - z$ close to the upper critical dimension~\cite{PhysRevLett.113.220401}. For a closed $\varphi^4$-model without an external bath, $\theta$ was recently derived in Refs.~\onlinecite{PhysRevB.91.220302,2015arXiv150604528M} using a different non-equilibrium renormalization-group method. In this paper, we complement our previous analysis by an expansion in $1/N$, where $N$ is the number of components of the field. We also largely extend our previous study, which focused on a quench starting from the symmetry-unbroken phase, by a direct investigation of the order parameter dynamics $\phi(t)$ starting from an initial state with non-zero $\phi$. We explicitly prove that the exponent $\theta$ is independent of the initial state and the particular quench protocol. All previous investigations were limited to space dimensions below the upper critical dimension. We also discuss the behavior at the upper critical dimension $d_{uc} = 4 - z$, where logarithmic corrections occur and 
\begin{align}
  \label{eq:12}
  \phi_{uc}(t) \propto \bigl[\ln(t/t_\gamma)\bigr]^{\theta_{uc} }
\end{align}
with $t_\gamma$ being a microscopic time-scale that marks the beginning of the universal pre-thermalization regime. 

The remainder of the paper is organized as follows: in Sec.~\ref{sec:model}, we introduce the model and the Hamiltonian, define the quench protocol and the coupling to the external bath. In Sec.~\ref{sec:non-equil-form}, we present a non-equilibrium formulation of the large-$N$ expansion and derive the large-$N$ saddle-point equations. To solve these equations self-consistently we employ a scaling form for the Green's functions that we discuss in Sec.~\ref{sec:scaling-behavior}. We present the large-N solution for a quench to the quantum critical point starting from the disorderd phase in Sec.~\ref{sec:quench-from-disord} and starting from the ordered phase in Sec.~\ref{sec:quench-start-order}. In Sec.~\ref{sec:light-cone-amplitude}, we relate the short-time scaling exponent $\theta$ to the time-dependence of the self-energy, which can be captured by a time-dependent effective mass, and determine the short-time behavior of the Green's functions. In Sec.~\ref{sc:long-time-GK}, we elaborate on the long-time behavior of the Green's function, which enters into the central calculation in Sec.~\ref{sec:self-consistent-solution} where we show the self-consistency of the solution, which fixes the value of the exponent $\theta$. In Sec.~\ref{sec:distribution-fct} we show that the system is characterized by a non-thermal distribution function at long-times. In Secs.~\ref{sec:upper-crit-dimens} and~\ref{sec:order-parameter-d_uc} we discuss the dynamics at the upper critical dimension. In Sec.~\ref{sec:quench-start-order}, we consider a quench that starts from the ordered phase, and find in Sec.~\ref{sec:preth-regime-cross} the scaling of the order parameter and the crossover time that separates the regimes of pre-thermal scaling dynamics from the adiabatic long-time dynamics that is characterized by equilibrium critical exponents as discussed in Sec.~\ref{sec:quasi-adiab-relax}. We finally summarize our main results and conclude in Sec.~\ref{sec:conclusions}. In the appendices, we provide further details on the analytical calculations, in particular on the different short-time scales in the problem in Appendix~\ref{sec:short-time-scales}, on the derivation of the large-N equations in Appendix~\ref{sec:large-n-expansion}, on the free post-quench Keldysh function in Appendix~\ref{app:post-quench G} and on the long-time limit of the full Green's functions in Appendix~\ref{app:long-time-limit}. 

\section{The model and the quench protocol}
\label{sec:model}
\subsection{The $\varphi^4$-model}
\label{sec:varphi4-model}
\begin{figure}[tb]
  \centering
  \includegraphics[width=\linewidth]{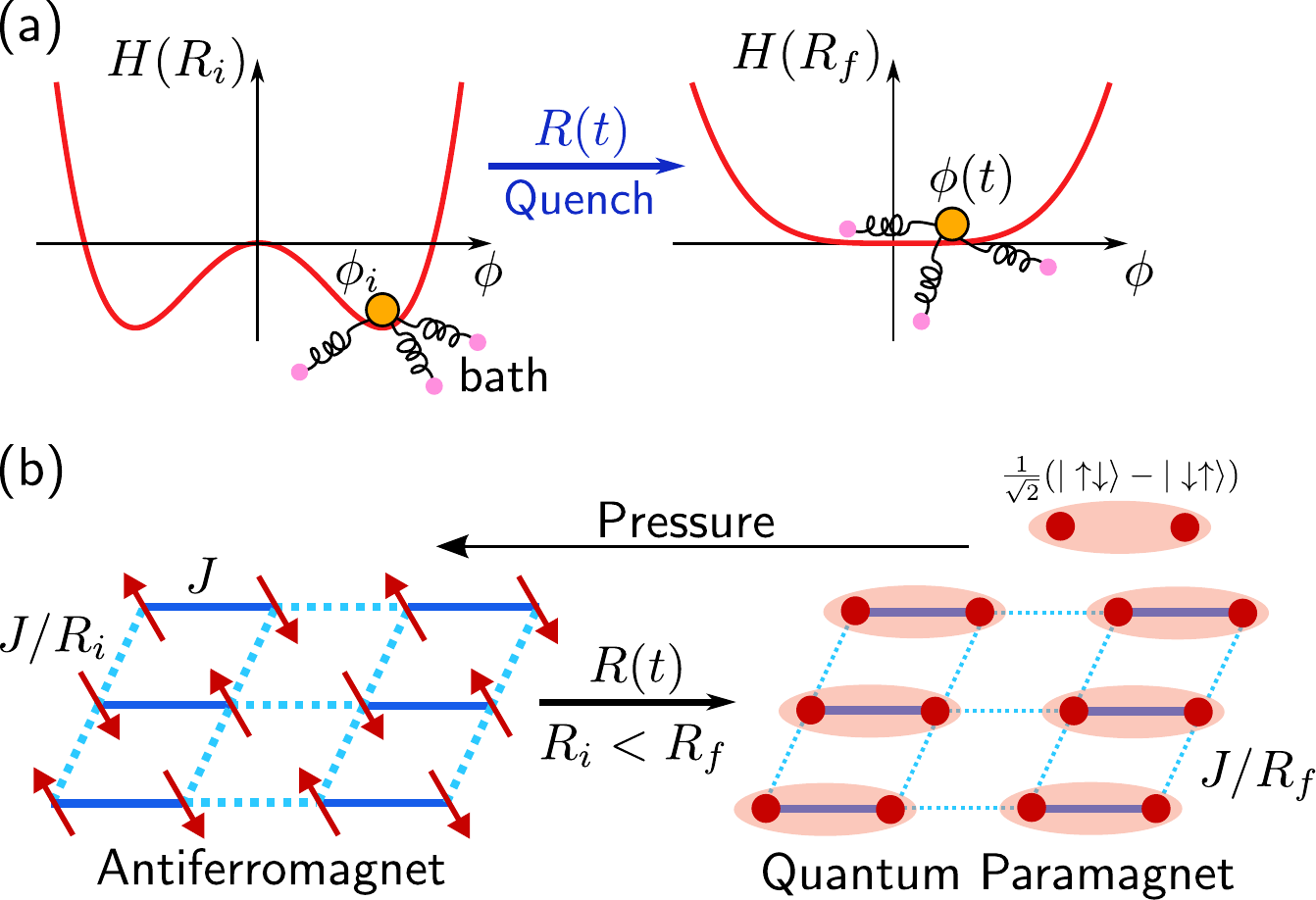}
  \caption{(Color online) (a) Schematic description of the $\varphi^4$-model coupled to an external bath. The order parameter $\phi$ experiences a potential landscape that depends on a set of parameters $R$. If the initial parameters $R_i$ are such that the system is prepared in the symmetry broken state with finite order parameter $\phi_i$, we consider a sudden change of the parameters from $R_i$ to $R_f$ that brings the system to the quantum critical point, where $\phi = 0$ in equilibrium. 
(b) One experimental realization of the $\varphi^4$ model for $N=3$ is a quantum dimer model in contact to phonons. The quench can be performed by quickly rapidly the pressure.  }
  \label{fig:1}
\end{figure}

We consider the post-quench non-equilibrium dynamics of a $\varphi^{4}$-model coupled to an external bath. The system is schematically depicted in Fig.~\ref{fig:1}(a). To be specific, we also show one particular experimental realization, the dimer antiferromagnet realized, for example, in $\text{Tl} \text{Cu} \text{Cl}_3$~\cite{PhysRevLett.100.205701}  in Fig.~\ref{fig:1}(b). Other realizations of our theory are discussed below. The $\varphi^{4}$-model is described by the Hamiltonian 
\begin{align}
H_{s}\left(t\right) = & \frac{1}{2}\int_{x}\left(\boldsymbol{\boldsymbol{\pi}}^{2}+\left(\nabla\boldsymbol{\boldsymbol{\varphi}}\right)^{2}+r_{0}\left(t\right)\boldsymbol{\varphi}^{2}\right.\nonumber \\
 + & \left.\frac{u\left(t\right)(\boldsymbol{\varphi}\cdot\boldsymbol{\varphi})^{2}}{2N}-\mathbf{h}\left(t\right)\cdot\boldsymbol{\boldsymbol{\varphi}}\right)\,.
\end{align}
Here, $\boldsymbol{\varphi}\left(x\right)=\left(\varphi_{1}\left(x\right),\dots,\varphi_{N}\left(x\right)\right)$
is an $N-$component scalar field and $\boldsymbol{\pi}\left(x\right)$
its canonically conjugated momentum with $\left[\varphi_{l}\left(x\right),\pi_{l'}\left(x'\right)\right]_{-}=i\delta_{ll'}\delta\left(x-x'\right)$.
In addition, we used the short hand notation $\int_{x}\equiv \int d^{d}x$.
The non-equilibrium dynamics is a consequence of the explicit time
dependence of either the bare mass $r_{0}\left(t\right)$, the interaction
strength $u\left(t\right)$, or the externally applied field $\mathbf{h}\left(t\right)$,
respectively. In the case of constant parameters the system is in equilibrium
and undergoes at temperature $T=0$ and for $\mathbf{h}=\mathbf{0}$
a quantum phase transition at a critical value $r_{0,c}\left(u\right)$
of the bare mass $r_{0}$. This transition can be captured analytically
by using an expansion for small $1/N$. In fact, the coefficient $1/N$
in front of the $\varphi^{4}-$term was introduced to allow for a
well defined $N\rightarrow\infty$ limit. The large-$N$ approximation
is complementary to our earlier non-equilibrium renormalization group
theory of Ref.~\onlinecite{PhysRevLett.113.220401}. In particular, it allows for a rather straightforward analysis
of the dynamics of a system that was initially in the symmetry broken
phase. 
\subsection{The quench protocol}
\label{sec:quench-protocol}
\begin{figure}[tb]
  \centering
  \includegraphics[width=.8\linewidth]{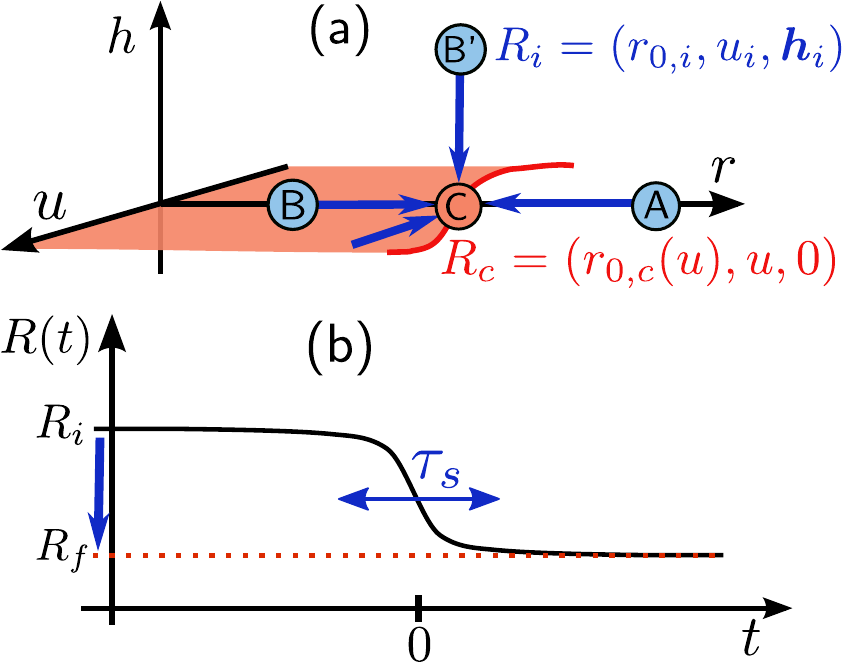}
  \caption{(Color online) (a) Schematic equilibrium phase diagram of the $\varphi^4$-model at $T=0$ including the different quench paths that we consider in this article. We either approach the quantum critical point $C$ from the symmetric side (path $A \rightarrow C$) or from the symmetry-broken side (path $B  \rightarrow C$). We also consider a quench in the presence of a magnetic field $\boldsymbol{h}_i$ that induces a finite initial order parameter $\phi_i$. (b) Schematic plot of the quench protocol. We consider fast quenches where the switching time $\tau_s$ from the initial parameter set $R_i$ to the final one $R_f$ occurs on a microscopic timescale.  }
  \label{fig:2}
\end{figure}

We consider the following protocols for the time dependence of 
\begin{equation}
R\left(t\right)\equiv\left(r_{0}(t),u\left(t\right),\mathbf{h}\left(t\right)\right).
\end{equation}
The system is initially prepared in the ground state with $R_{i}=\left(r_{0,i},u_{i},\mathbf{h}_{i}\right)$ somewhat away from the critical point located at $R_{c}=\left(r_{0,c}\left(u\right),u,0\right)$. 
Then we switch $R(t)$ such that it approaches a final value $R(t\rightarrow\infty)=R_{f}$ that is closer to the critical point. The schematic phase diagram showing the parameter quenches that we consider in the following are shown in Fig.~\ref{fig:2}(a). The switching of $R(t)$ takes place on a timescale $\tau_{s}$. We consider the limit where $\tau_{s}$ is of the order of the microscopic timescales of the system, i.e. the fastest time that is consistent with the validity of the $\varphi^{4}-$model
as an effective low energy description. This situation is depicted in Fig.~\ref{fig:2}(b). Keeping this in mind, we can
safely perform the limit $\tau_{s}\rightarrow0$ in the calculation
such that 
\begin{equation}
R(t)=R_{i}+\theta\left(t\right)\left(R_{f}-R_{i}\right),\label{eq:r switch}
\end{equation}
with step-function $\theta\left(t\right)$. 

After the quench the system instantly falls out of equilibrium. In
the context of condensed matter systems, the generic situation is
an open system that is coupled to external bath degrees of freedom.
The bath, here conveniently described by a set of harmonic oscillators,
is assumed to stay at a constant temperature, $T$. In particular,
if $T=0$ and $R_{f}=R_{c}$, the system does indeed reach the quantum
critical point in the limit $t\rightarrow\infty$. This is to be contrasted
with the behavior of a closed system, where energy conservation alone
implies that the system will not be in the ground state of the post-quench
Hamiltonian~\cite{PhysRevB.81.134305,PhysRevB.91.220302}. Even if the system thermalizes, a quench towards the critical point will heat up a closed system to a finite temperature. 
\subsection{Coupling to an external bath}
\label{sec:coupling-an-external}
If we include the coupling to the bath, the full Hamiltonian reads
\begin{equation}
H(t)=H_{s}(t)+H_{b}+H_{sb}.\label{eq:initial H}
\end{equation}
Here, $H_{b}$ describes the bath of harmonic oscillators and $H_{sb}$
the linear coupling between system and bath: 
\begin{align}
H_{b}= & \frac{1}{2}\sum_{j}\int d^{d}x(\mathbf{P}_{j}^{2}+\Omega_{j}^{2}\mathbf{X}_{j}^{2}),\nonumber \\
H_{sb}= & \sum_{j}c_{j}\int d^{d}x\mathbf{X}_{j}\cdot\boldsymbol{\varphi}.\label{eq:bath coupl}
\end{align}
In accordance with Eq.~\eqref{eq:r switch} the full Hamiltonian before
and after the quench is then $H_{i}=H\left(t<0\right)$ and $H_{f}=H\left(t>0\right)$,
respectively.

In Eq.~\eqref{eq:bath coupl} $\Omega_{j}$ are the frequencies of bath
oscillators that couple with coupling constants $c_{j}$. As the bath
stays in equilibrium throughout, it is fully characterized in terms
of the retarded$ $ function 
\begin{equation}
\eta(\omega)=-\sum_{j}\frac{c_{j}^{2}}{(\omega+i0^{+})^{2}-\Omega_{j}^{2}}.\label{eq:eta(omega)}
\end{equation}
The imaginary part of $\eta\left(\omega\right)$ is the spectral
function of the bath. We consider a bath with spectral function 
\begin{equation}
{\rm Im}\eta(\omega)=\gamma\omega|\omega|^{\alpha-1}e^{-|\omega|/\omega_{c}}.
\end{equation}
Here, the damping coefficient $\gamma$ determines the overall strength
of the coupling to the bath and $\omega_{c}$ is the ultra-violet
cutoff of the bath spectrum. The exponent $\alpha$ characterizes
the low-frequency behavior. $\alpha=1$ corresponds to an ohmic, $\alpha>1$
to super-ohmic and $\alpha<1$ to a sub-ohmic bath, respectively~\cite{weissdissipation}.
The frequency dependence of Im$\eta\left(\omega\right)$ for different
$\alpha$ is depicted in Fig.~\ref{fig:KeldyshContour}(b). A Kramers-Kronig transformation yields
the real part of $\eta\left(\omega\right)$. We obtain 
\begin{eqnarray}
\delta\eta\left(\omega\right) & \equiv & \eta\left(\omega\right)-\eta\left(0\right)\nonumber \\
 & = & \gamma\left(-\cot\left(\frac{\pi\alpha}{2}\right)+i{\rm sign}\left(\omega\right)\right)\left\vert \omega\right\vert ^{\alpha}\label{eq:deta}
\end{eqnarray}
for frequencies small in magnitude compared to the cut off $\omega_{c}$.
The zero-frequency value $\eta\left(0\right)=\frac{\gamma\alpha}{2\pi}\omega_{c}^{\alpha}$
depends explicitly on the value of the bath cut off. As we will see,
$\eta\left(0\right)$ merely shifts the location of the quantum critical
point of the system but does not affect the generic behavior near
it. Below we will also need the analytic continuation $\eta^{M}\left(\omega_{n}\right)=\eta\left(0\right)+\delta\eta^{M}\left(\omega_{n}\right)$
of $\eta\left(\omega\right)$ to the Matsubara axis ($\omega+i0^{+}\rightarrow i\omega_{n}=i2n\pi T$).
It holds 
\begin{equation}
\delta\eta^{M}\left(\omega_{n}\right)=-\frac{\gamma}{\sin\frac{\pi\alpha}{2}}\left|\omega_{n}\right|^{\alpha},\label{eq:etaMatsubara}
\end{equation}
again valid for frequencies that are small compared to the bath cut off: $\left|\omega_{n}\right|\ll\omega_{c}$.
\subsection{Experimental realizations}
\label{sec:exper-real}
There are a number of experimental systems that can be effectively described by an $N$-component $\varphi^4$-theory that we are considering. Realizations for $N=1$ are the magnetic insulators $\text{Co} \text{Nb}_2 \text{O}_6$~\cite{Coldea08012010} and $\text{Li} \text{Ho}_x \text{Y}_{1− x} \text{F}_4$~\cite{Brooke30041999}, which can be described by the transverse field quantum Ising model. Dissipation in these systems arises via coupling to phonons. Realizations for $N=2$ are systems near the superconductor-insulator (or superfluid-insulator) transition, Josephson junction arrays and quantum antiferromagnets in a magnetic field~\cite{sachdev_qpt_book}. Superfluid-insulator transitions in the Bose-Hubbard model have been experimentally investigated using cold-atom quantum gases confined to optical lattice potentials~\cite{greiner_nature_2002,bakr_probing_2010}. Dissipation in those system has been engineered, for example, by coupling to other species~\cite{PhysRevLett.105.045303}. Dissipative nanowires near a transition to a superconducting state~\cite{sachdev_universal_2004} and an ensemble of qubits in a photon cavity~\cite{koch:023811, HouchTuereciKoch-NatPhys-2012} provide further realizations for $N=2$. A realization for $N=3$ is the quantum dimer antiferromagnet $\text{Tl} \text{Cu} \text{Cl}_3$ shown in Fig.~\ref{fig:1}(b), which can be driven across the quantum phase transition by either changing pressure~~\cite{PhysRevLett.100.205701} or magnetic field~\cite{Ruegg-Nature-2003}.

\section{Non-equilibrium formulation of the large-$N$ expansion}
\label{sec:non-equil-form}
The natural technique to treat the post-quench dynamics is the non-equilibrium many body formalism due to Schwinger~\cite{Schwinger-NonEqSchwingerKeldysh-1961}, Kadanoff-Baym~\cite{KadanoffBaym-QuantumStatisticalMechanicsBook-1962}, and Keldysh~\cite{Keldysh-NonEqSchwingerKeldysh-1965} (see also Ref.~\onlinecite{Kamenev-NonEqFieldTheory-Book}), where the field theory is placed on the three-time Keldysh contour shown in Fig.~\ref{fig:KeldyshContour}(a). A subtlety of our problem is that the initial state at $t=0$ is itself a many-body state where (i) bath and system are entangled and (ii) interactions cannot be neglected. Thus, we cannot assume that we evolve from a non-interacting initial state and switch on the interactions adiabatically afterwards. How to modify the approach to the scenario where we prepare system and bath in an entangled interacting equilibrium state governed by
$H_{i}$ and where the subsequent time evolution is then determined
by $H_{f}$ was discussed by Danielewicz~\cite{Danielewicz-AnnPhys-1984} and Wagner~\cite{PhysRevB.44.6104}. In what follows we will first summarize and then use this approach.

\subsection{Action within the three-time-contour formalism}
\label{sec:action-within-three}
\begin{figure}[tb]
  \centering
  \includegraphics[width=\linewidth]{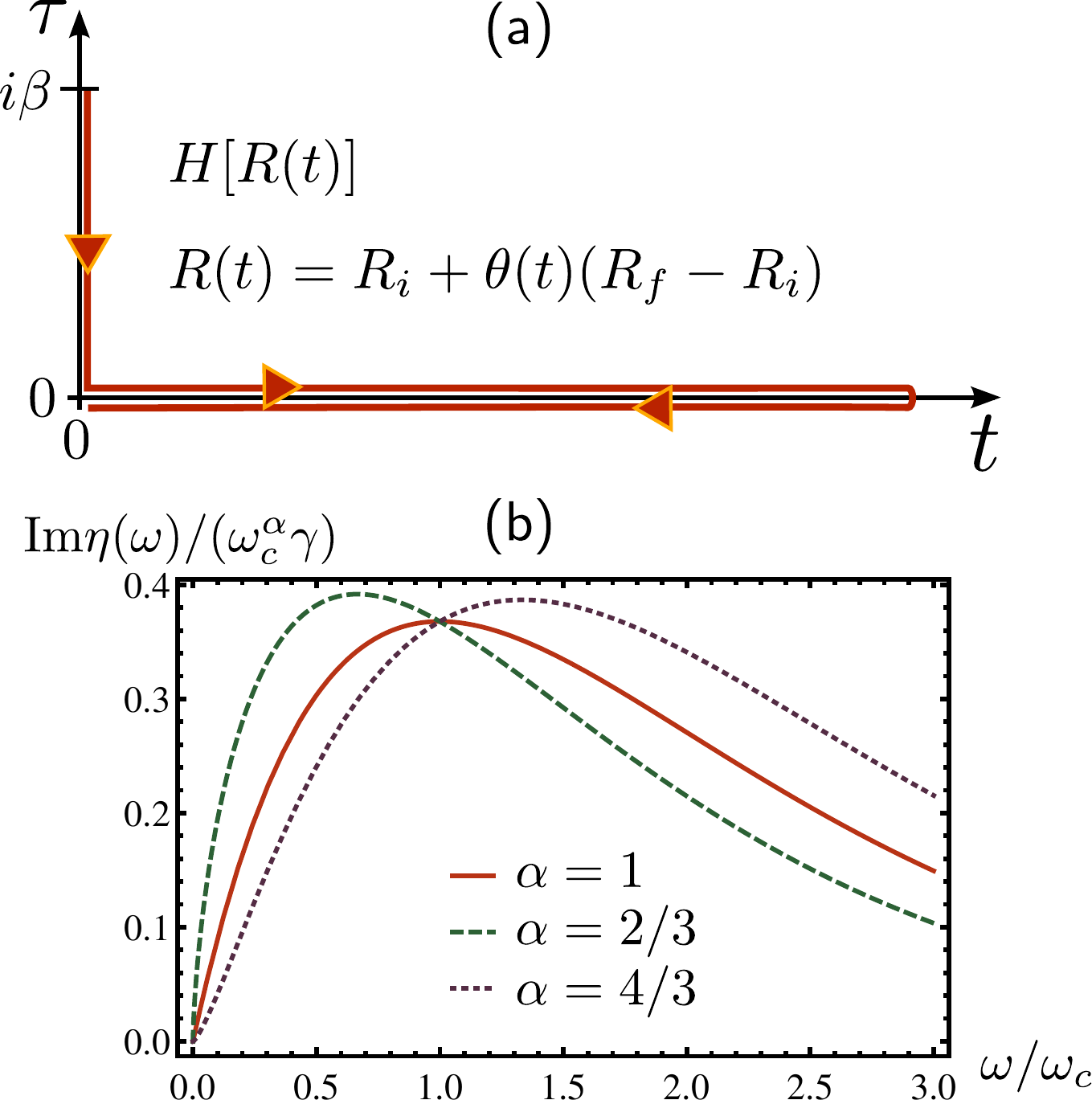}
  \caption{(Color online) (a) Keldysh three-time contour. (b) Bath spectral function for different power law exponents corresponding to the sub-ohmic case $\alpha < 1$, the ohmic case $\alpha = 1$ and the super-ohmic case $\alpha > 1$.  }
  \label{fig:KeldyshContour}
\end{figure}

Let us consider an initial state of system and bath in equilibrium
at temperature $T$. The state right before the quench at $t=0$ is
then characterized by the density matrix 
\begin{equation}
\rho_{i}=\frac{1}{Z}e^{-\beta H_{i}},
\end{equation}
where $\beta=1/T$ and $Z={\rm tr}e^{-\beta H_{i}}$ is the pre-quench
equilibrium partition function. In the limit $T\rightarrow0$ it holds
$\rho_{i}=\left|\Psi_{0}\right\rangle \left\langle \Psi_{0}\right|$,
where $\left|\Psi_{0}\right\rangle $ is the ground state of the coupled
system and bath with Hamiltonian $H_{i}$. The subsequent time evolution
is governed by $H_{f}$, i.e. the density matrix is given as 
\begin{equation}
\rho\left(t\right)=U\left(t,0\right)\rho_{i}U\left(0,t\right),
\end{equation}
where $U\left(t,t'\right)=\exp\left(-i\left(t-t'\right)H_{f}\right).$
The expectation value of an arbitrary operator (for specificity we
consider our scalar field $\boldsymbol{\varphi}$) is then 
\begin{eqnarray}
\left\langle \varphi\left(t\right)\right\rangle  & = & {\rm tr}\left(\rho\left(t\right)\varphi\right)\nonumber \\
 & = & \frac{1}{Z}{\rm tr}\left(e^{-\beta H_{i}}U\left(0,t\right)\varphi U\left(t,0\right)\right).
\end{eqnarray}
For simplicity we suppress for the moment the component index $l$
and spatial coordinate $x$ of the field, i.e. $\varphi\left(t\right)=\varphi_{l}\left(x,t\right)$.
The time evolution that enters the expectation value can be efficiently
placed on the contour ${\cal C}$ shown in Fig.~\ref{fig:KeldyshContour}(a) with evolution
operator such that 
\begin{equation}
\left\langle \varphi\left(t\right)\right\rangle =\frac{{\rm tr}\left(T_{{\cal C}}e^{-i\int_{{\cal C}}dsH\left(s\right)}\varphi\left(t\right)\right)}{{\rm tr}\left(T_{{\cal C}}e^{-i\int_{{\cal C}}dsH\left(s\right)}\right)}.
\end{equation}
The time argument of $\varphi$ denotes that the Schr\"odinger operator
has to be inserted at time $t$. $T_{{\cal C}}$ is the time ordering
operator along the contour from $i\beta\rightarrow0\rightarrow\infty\rightarrow0$.
In the denominator we used the fact that $e^{-\beta H_{i}}=U\left(0,t\right)U\left(t,0\right)e^{-\beta H_{i}}$.$ $
In full analogy one can consider two-time correlation functions 
\begin{eqnarray}
{\cal G}\left(t,t'\right) & = & -i\left\langle T_{{\cal C}}\varphi\left(t\right)\varphi\left(t'\right)\right\rangle \nonumber \\
 & = & -i\frac{1}{Z}{\rm tr}\left(T_{{\cal C}}e^{-i\int_{{\cal C}}dsH\left(s\right)}\varphi\left(t\right)\varphi\left(t'\right)\right).
\end{eqnarray}
To determine these and other observables we consider the generating
functional on the contour ${\cal C}$ ($\int_{t\in{\cal C}}\equiv\int_{{\cal C}}dt$)
\begin{equation}
W\left[\mathbf{h}\right]=\int D\boldsymbol{\varphi}D\mathbf{X}e^{iS\left[\boldsymbol{\varphi},\mathbf{X}\right]-i\int_{x,t\in{\cal C}}\mathbf{h}\left(t\right)\cdot\boldsymbol{\varphi}\left(t\right)}
\end{equation}
with action 
\begin{equation}
S\left[\boldsymbol{\varphi},\mathbf{X}\right]=S_{s}\left[\boldsymbol{\varphi}\right]+S_{b}\left[\mathbf{X}\right]+S_{sb}\left[\boldsymbol{\varphi},\mathbf{X}\right].
\end{equation}
$S\left[\boldsymbol{\varphi},\mathbf{X}\right]$ consists of the action
of the system, the bath, and the coupling term. The individual terms
are 
\begin{eqnarray}
S_{s} & = & \frac{1}{2}\int_{x,t\in{\cal C}}\left\{ \left(\partial_{t}\boldsymbol{\varphi}\left(t\right)\right)^{2}-r_{0}\left(t\right)\boldsymbol{\varphi}\left(t\right)^{2}\right.\nonumber \\
 &  & -\left(\nabla\boldsymbol{\varphi}\left(t\right)\right)^{2}-\left.\frac{u\left(t\right)}{2N}\boldsymbol{\varphi}\left(t\right)^{4}\right\} \label{eq:action_system}
\end{eqnarray}
for the action of the system and 
\begin{eqnarray}
\label{eq:13}
S_{b} & = & \frac{1}{2}\sum_{j}\int_{x,t\in{\cal C}}\left(\left(\partial_{t}\boldsymbol{X}_{j}\left(t\right)\right)^{2}-\Omega_{j}^{2}\boldsymbol{X}_{j}\left(t\right)^{2}\right) \\
\label{eq:14}
S_{sb} & =- & \sum_{j}c_{j}\int_{x,t\in{\cal C}}\mathbf{X}_{j}\left(t\right)\cdot\boldsymbol{\varphi}\left(t\right)
\end{eqnarray}
for the bath action and the system-bath coupling, respectively.

We integrate out the bath variables at the expense of a bare propagator
that is highly non-local in time: 
\begin{eqnarray}
{\cal G}_{0}^{-1}\left(t,t'\right) & = & -\left(\partial_{t}^{2}+r_{0}\left(t\right)-\nabla^{2}\right)\delta\left(t-t'\right)\nonumber \\
 & + & \Delta\left(t-t'\right).
\end{eqnarray}
The effects of the bath enter via the non-local self energy $\Delta\left(t-t'\right)$
that is formally given as 
\begin{equation}
\Delta\left(t-t'\right)=-\sum_{j}c_{j}^{2}\left(\partial_{t}^{2}+\Omega_{j}^{2}\right)^{-1}\delta\left(t-t'\right).
\end{equation}
The resulting action 
\begin{eqnarray}
S\left[\boldsymbol{\varphi}\right] & = & \frac{1}{2}\int_{x,t, t'\in{\cal C}}\boldsymbol{\varphi}\left(t\right){\cal G}_{0}^{-1}\left(t,t'\right)\boldsymbol{\varphi}\left(t'\right)\nonumber \\
 & - & \frac{1}{4N}\int_{x,t\in{\cal C}}u\left(t\right)\left(\boldsymbol{\varphi}\left(t\right)\cdot\boldsymbol{\varphi}\left(t\right)\right)^{2}. \label{eq:23}
\end{eqnarray}
depends only on the collective field $\boldsymbol{\varphi}$ and yields the generating functional via 
\begin{equation}
W\left[\mathbf{h}\right]=\int D\boldsymbol{\varphi}e^{iS\left[\boldsymbol{\varphi}\right]-i\int_{x,t\in{\cal C}}\mathbf{h}\left(t\right)\cdot\boldsymbol{\varphi}\left(t\right)}.
\end{equation}

In the usual Schwinger-Keldysh formalism it is convenient to place 
the four possible arrangements of the times $t$ and $t'$ on the
forward and backward branch of the contour in a $2\times2$ matrix.
In our case we have three branches of the contour which can be captured
in terms of a $3\times3$ matrix structure of the Green's function \cite{PhysRevB.44.6104}
\begin{equation}
{\cal G}=\left(\begin{array}{ccc}
iG^{M} & \tilde{G}^{<} & \tilde{G}^{<}\\
\tilde{G}^{>} & G^{T} & G^{<}\\
\tilde{G}^{>} & G^{>} & G^{\bar{T}}
\end{array}\right).
\end{equation}
The first matrix-element is up to a factor $i$ the imaginary time
Green's function in equilibrium prior to the quench: 
\begin{equation}
G^{M}\left(\tau-\tau'\right)=-\left\langle T_{\tau}\varphi_{M}\left(\tau\right)\varphi_{M}\left(\tau'\right)\right\rangle ,
\end{equation}
where $\varphi_{M}\left(\tau\right)=e^{\tau H_{i}}\varphi e^{-\tau H_{i}}$.
$T_{\tau}$ is the time ordering operator along the vertical segment
of the contour. The inverse of the bare Matsubara function is 
\begin{eqnarray}
(g_{i}^{M})^{-1}\left(\tau,\tau'\right) & = & \left(\partial_{\tau}^{2}-r_{0,i}+\nabla^{2}\right)\delta\left(\tau-\tau'\right)\nonumber \\
 & + & \eta^{M}\left(\tau-\tau'\right),
\end{eqnarray}
where $\eta^{M}\left(\tau\right)$ is the Fourier transform of $\eta^{M}\left(\omega_{n}\right)$
in Eq.~\eqref{eq:etaMatsubara}. $G^{M}(\tau, \tau')\equiv G^{M}(\tau- \tau')$ only depends on the difference
between the two time variables. The Fourier transform of the bare Matsubara function yields
\begin{equation}
g_{i}^{M}\left(\omega_{n}\right)=\frac{1}{-\omega_{n}^{2}-\bar{r}_{0,i}-q^{2}+ \delta \eta^{M}\left(\omega_{n}\right)}.\label{eq:G0Matsubara}
\end{equation}
Here, we have introduced $\bar{r}_{0,i} = r_{0,i} - \eta^M(0)$ to account for a trivial shift of the bare mass due to the bath coupling. 
The functions 
\begin{eqnarray}
\tilde{G}^{<}\left(\tau,t'\right) & = & i\left\langle \varphi_{M}\left(\tau\right)\varphi_{H}\left(t'\right)\right\rangle \nonumber \\
\tilde{G}^{>}\left(\tau,t'\right) & = & i\left\langle \varphi_{H}\left(t'\right)\varphi_{M}\left(\tau\right)\right\rangle 
\end{eqnarray}
describe correlations across the quench, where $\varphi_{H}\left(t\right)=e^{iH_{f}t}\varphi e^{-iH_{f}t}$
is given in the Heisenberg representation after the quench. The remaining
$2\times2$ block with 
\begin{eqnarray}
G^{T}\left(t,t'\right) & = & -i\left\langle T_{t}\varphi_{H}\left(t\right)\varphi_{H}\left(t'\right)\right\rangle ,\nonumber \\
G^{\tilde{T}}\left(t,t'\right) & = & -i\left\langle \tilde{T}_{t}\varphi_{H}\left(t\right)\varphi_{H}\left(t'\right)\right\rangle ,\nonumber \\
G^{>}\left(t,t'\right) & = & -i\left\langle \varphi_{H}\left(t'\right)\varphi_{H}\left(t\right)\right\rangle ,\nonumber \\
G^{<}\left(t,t'\right) & = & -i\left\langle \varphi_{H}\left(t\right)\varphi_{H}\left(t'\right)\right\rangle 
\end{eqnarray}
makes up the usual Keldysh matrix where $T_{t}$ and $\tilde{T_{t}}$
refer to the time ordering and anti-time ordering operator along the horizontal real-time branch of the contour $\mathcal{C}$, respectively.
Of those 9 Green's functions that occur in the $3\times3$ matrix
structure, only a few contain truly independent information. By transforming
${\cal G}$ with the rotation matrix 
\begin{equation}
L=\left(\begin{array}{ccc}
-1 & 0 & 0\\
0 & \frac{1}{\sqrt{2}} & \frac{1}{\sqrt{2}}\\
0 & \frac{1}{\sqrt{2}} & -\frac{1}{\sqrt{2}}
\end{array}\right),
\end{equation}
one can benefit from this redundancy among the Green's functions and
obtain
\begin{equation}
L^{-1}{\cal G}L=\left(\begin{array}{ccc}
G^{M} & -\sqrt{2}\tilde{G}^{<} & 0\\
-\sqrt{2}\tilde{G}^{>} & G^{K} & G^{R}\\
0 & G^{A} & 0
\end{array}\right).\label{eq:G-matrix}
\end{equation}
The scalar field forms a three component vector $\left(\varphi_{M},\varphi_{H}^{c},\varphi_{H}^{q}\right)$with
so called classical and quantum components 
\begin{equation}
\varphi^{c,q}_{H}\left(t\right)=\frac{1}{\sqrt{2}}\left(\varphi_{H}^{+}\left(t\right)\pm\varphi_{H}^{-}\left(t\right)\right),
\end{equation}
in addition to the Matsubara component.
The two relevant Green's functions on the forward and backward branch
of the contour are the retarded Green's function 
\begin{equation}
G^{R}\left(t,t'\right)=-i\theta\left(t-t'\right)\left\langle \left[\varphi_{H}\left(t\right),\varphi_{H}\left(t'\right)\right]_{-}\right\rangle \label{eq:defGR}
\end{equation}
and the Keldysh function 
\begin{equation}
G^{K}\left(t,t'\right)=-i\left\langle \left[\varphi_{H}\left(t\right),\varphi_{H}\left(t'\right)\right]_{+}\right\rangle \,,
\label{eq:defGK}
\end{equation}
where $[A,B]_{\pm} = AB \pm BA$. The retarded function measures the response of the order parameter
at time $t$ caused by an external field that couples to it and that
acted at time $t'<t$ after the quench, while the advanced function
\begin{equation}
G^{A}\left(t,t'\right)=i\theta\left(t'-t\right)\left\langle \left[\varphi_{H}\left(t\right),\varphi_{H}\left(t'\right)\right]_{-}\right\rangle \label{eq:defGA}
\end{equation}
contains the same information as the retarded function: $G^{A}\left(t,t'\right)=G^{R}\left(t',t\right)$.
In distinction, the Keldysh function is a measure of the strength
of correlations between order parameter configurations at distinct
times after the quench. Couplings between the horizontal and vertical
branches of the contour are determined by correlation functions $\tilde{G}^{<}$
and $\tilde{G}^{>}$. These functions take into account the coupling
``across the quench''. Notice, that the Matsubara field $\varphi_{M}$
only couples to the quantum component $\varphi^{q}_H$. For the bare Green's functions this implies that only the Keldysh function contains pre-quench memory effects. Thus, the bare post-quench Keldysh function $g_{f}^{K}\left(t,t'\right)$ of a non-interacting system will depend on both time variables, i.e., display aging effects. How
to determine $g_{f}^{K}\left(t,t'\right)$ will be discussed in detail below. The bare retarded function only depends on the time differences
$g_{f}^{R}\left(t,t'\right)=g_{f}^{R}\left(t-t'\right)$ (both before and after the quench) and can be Fourier transformed to yield
\begin{equation}
\label{eq:Goretarded}
g_{f}^{R}(\omega)=\frac{1}{\omega^{2}-\bar{r}_{0,f} - q^{2} + \delta \eta(\omega)}\,,
\end{equation}
where $\bar{r}_{0,f} = r_{0,f} - \eta(0)$. Notice, that this form is only correct for the bare retarded function, i.e. for $u=0$. As we will see, many-body interactions couple response and correlation functions and lead to aging effects in the retarded function as well. 

\subsection{Large-N equations}
\label{sec:large-n-equations}
We consider interaction effects within the large-$N$ approach. For a large number of field components $N$ the generating functional $W[\boldsymbol{h}]$ can be evaluated in the saddle-point approximation controlled in small $1/N$. In Appendix~\ref{sec:large-n-expansion}, we explicitly perform this approximation on the three-time-contour and summarize the main steps of the large-$N$ analysis for a system with quench. 
The resulting self-consistent large-$N$ equations before the quench read
\begin{eqnarray}
\label{eq:19}
h_{i} & = & r_{i}\phi_{i} \\
\label{eq:20}
r_{i} & = & \bar{r}_{0,i}+\frac{u_{i}}{2}\phi_{i}^{2}+u_{i}\int_{q,\omega_{n}}G_{r_i}^{M}\left(q,\omega_{n}\right)
\end{eqnarray}
and the equations after the quench read
\begin{eqnarray}
h_{f} & = & -\int_{0}^{\infty}dt'(G_{r}^{R})^{-1}\left(t,t'\right)\phi\left(t'\right) \nonumber \\
& &-\phi_{i}\int_{-\infty}^{0}dt'\delta\eta\left(t-t'\right) \label{eq:eomlN} \\
r\left(t\right) & = & \bar{r}_{0,f}+\frac{u_{f}}{2}\phi^{2}\left(t\right)+\frac{u_f}{2}\int_{q} iG_{r}^{K}\left(q,t,t\right).\label{eq:largeN-time-dep-mass}
\end{eqnarray}
We use the notation $\int_{q}\equiv\int\frac{d^{d}q}{\left(2\pi\right)^{d}}$
and $\int_{\omega_{n}}\equiv T\sum_{n}$ for the momentum integration
and the summation over Matsubara frequencies. $\phi_{i}$ and $r_{i}$
are the order parameter and renormalized mass prior to the quench. $G_{r}^{M}$ is the
renormalized Matsubara function, where $\bar{r}_{0,i}$ in Eq.~\eqref{eq:G0Matsubara} is replaced by $r_{i}$. In full analogy
is the renormalized retarded function 
\begin{eqnarray}
(G_{r}^{R})^{-1}\left(t,t'\right) & = & -\left(\partial_{t}^{2}+r\left(t\right)-\nabla^{2}\right)\delta\left(t-t'\right)\nonumber \\
 & + & \delta\eta\left(t-t'\right)\label{eq:GRrenorm}
\end{eqnarray}
 governed by the renormalized, time-dependent mass $r\left(t\right)$. The non-equilibrium
dynamics leads to a time-dependence of the mass such that the response
function is affected by aging behavior. The set of equations is closed
by the Keldysh function that can be expressed in the form
\begin{equation}
G_{r}^{K}(t,t')=\int_{s,s'}G_{r}^{R}(t,s)M(s,s')G_{r}^{A}(s,t')\,,
\label{eq:GKrenorm}
\end{equation}
where we introduced the memory function $M\left(s,s'\right)$ that is explicitly determined in Appendix \ref{app:post-quench G}. Once this memory function is known we have a closed set of equations for the non-equilibrium dynamics after a quantum quench. 

For our subsequent analysis it is convenient to rewrite the last two equations in terms of an expansion in 
\begin{align}
  \label{eq:6}
  \delta r\left(t\right)\equiv r\left(t\right)-r_f \,,
\end{align}
where $r_{f}\equiv r\left(t\rightarrow\infty\right)$ is the equilibrium value of the renormalized mass for the post quench parameters (see Eq.~\eqref{eq:eq1} below). This includes a shift due to interactions. It implies in particular that $r_{f}=0$ for a quench right to the critical point. The renormalized propagators can then be obtained from the Dyson equations
\begin{eqnarray}
G_{r}^{R}\left(t,t^{\prime}\right) & = & g^{R}\left(t-t^{\prime}\right)\label{eq:DysonR}\\
 & + & \int_{t'}^{t}dsg^{R}\left(t-s\right)\delta r\left(s\right)G_{r}^{R}\left(s,t^{\prime}\right)\nonumber 
\end{eqnarray}
and 
\begin{eqnarray}
G_{r}^{K}\left(t,t'\right) & = & g^{K}\left(t,t'\right)\label{eq:DysonK}\\
 & + & \int_{0}^{t'}dsg^{K}\left(t,s\right)\delta r\left(s\right)G_{r}^{A}\left(s,t'\right)\nonumber \\
 & + & \int_{0}^{t}dsg^{R}\left(t-s\right)\delta r\left(s\right)G_{r}^{K}\left(s,t'\right).\nonumber  \,.
\end{eqnarray}
Here, the function $g^{R}\left(t-t^{\prime}\right)$ is the solution of Eq.~\eqref{eq:GRrenorm} with $r\left(t\right)$ replaced by $r_{f}$, i.e. it reads in Fourier space as 
\begin{align}
  \label{eq:7}
  g^R(\omega) = \frac{1}{\omega^2 - r_f - q^2 + \delta \eta(\omega)}\,.
\end{align}
The Keldysh function $g^K$ is determined as in Eq.~\eqref{eq:GKrenorm} via $g^{K}\left(t,t'\right) = \int_{s,s'} g^R(t-s) M(s,s') g^A(s'-t')$.

In Secs.~\ref{sec:quench-from-disord} and~\ref{sec:quench-start-order} we present solutions of the large-$N$ equations for several different quench protocols, both starting in the disorderd and in the ordered phase. In the next sections, however, we first summarize the equilibrium limit without quench and motivate the scaling behavior out-of-equilibrium of various observables that will be confirmed by the subsequent explicit calculation.

\subsection{The equilibrium limit of the large-N analysis}
\label{sec:equil-limit-large}
Without quench, the system remains in equilibrium and the large-$N$
equations before and after the quench become identical. In equilibrium,
the Keldysh function $G^{K}(q,t,t')$ only depends on the time difference
$t-t'$, with Fourier transform determined by the fluctuation-dissipation
theorem: 
\begin{equation}
G_{r}^{K}\left(q,\omega\right)=2i\coth\left(\frac{\omega}{2T}\right){\rm Im}G_{r}^{R}\left(q,\omega\right).
\end{equation}
Using 
\begin{equation}
T\sum_{n}G_{r}^{M}\left(q,\omega_{n}\right)=\frac{1}{2}\int_{-\infty}^{\infty}\frac{d\omega}{2\pi}iG_{r}^{K}\left(q,\omega\right)
\end{equation}
it follows with $\int_{-\infty}^{\infty}dt\delta\eta\left(t\right)=0$
that the equations before and after the quench give, as expected,
the same solution. We can drop the subscripts $i$ and $f$ that
distinguish between the pre- and post-quench regimes. The renormalized mass fulfills
\begin{eqnarray}
r & = & \bar{r}_{0}+u\int_{q,\omega}G_{r}^{M}(q,\omega)+ \frac u2\phi^{2}\,.
\label{eq:eq1}
\end{eqnarray}
The equation of state for the order parameter reads 
\begin{equation}
r\phi=h.\label{eq:eq2}
\end{equation}
In the ordered phase it holds that $\phi\neq0$ and using Eq.~\eqref{eq:eq2} it thus follows for $h=0$ that $r\left(h=0\right)=0$. The
excitation spectrum is massless due to the Goldstone theorem and the
fact that longitudinal excitations are of order $1/N$. The transition
takes place when $r=0$ and $\phi=0$ and determines the critical
value $\bar{r}_{0,c}\left(u\right)$. For $T=0$ the latter is
determined by
\begin{equation}
\bar{r}_{0,c}=-u\int_{q,\omega}\frac{1}{\omega^{2}+q^{2}+\delta\eta^{M}\left(\omega\right)},\label{eq:r0equil}
\end{equation}
$\bar{r}_{0,c}$ depends on the momentum cut off $\Lambda$ and, via
$\eta\left(0\right)$, on the frequency cut off of the bath spectrum.
However, universal behavior emerges as a function of the distance $\delta r=\bar{r}_{0} - \bar{r}_{0,c}$
to the quantum critical point. For sufficiently small energies the
dynamics is dominated by the bath. Comparing length and time scales
in the propagator with $\delta \eta^M(\omega) \propto |\omega|^\alpha$ yields the dynamic scaling exponent 
\begin{equation}
\label{eq:15}
z=2/\alpha.
\end{equation}
Analyzing Eqs.~\eqref{eq:eq1} and~\eqref{eq:eq2} one obtains below the upper critical dimension $d<d_{uc}=4-z$ the well known results for the order parameter $\boldsymbol{\left|\phi\right|}\propto(-\delta r)^{\beta}$
with $\beta=1/2$, for the correlation length $\xi=r^{-1/2}\propto \delta r^{-\nu}$
with $\nu^{-1}=d+z-2$, and for the order-parameter susceptibility at
the critical point $G^{R}\left(q=0,\omega=0\right)\propto \delta r^{-\gamma}$
with $\gamma=2\nu.$ For $d>d_{uc}$ all exponents take their mean
field values.

\section{Scaling behavior}
\label{sec:scaling-behavior}
In this section we discuss scaling arguments for the non-equilibrium
dynamics after a quench towards the critical point. In addition to
the dynamics of the order parameter $\phi(t) \hat{\boldsymbol{\varphi}} =\left\langle \boldsymbol{\varphi}\left(k,t\right)\right\rangle $
we are interested in the retarded and Keldysh Green's functions $G^{R}\left(k,t,t'\right)$
and $G^{K}\left(k,t,t'\right)$, respectively, where both time arguments
are after the quench. We suppress the field index $l$ here and in the following for simplicity. These scaling relations and the corresponding
exponents will be determined from a self-consistent solution of the
coupled large-$N$ equation in the subsequent sections.

In equilibrium, the order-parameter at $T=0$ obeys 
\begin{equation}
\phi_{{\rm eq}}\left(\delta r,h\right)=b^{-\beta/\nu}\phi_{{\rm eq}}\left(b^{1/\nu}\delta r,b^{\beta\delta/\nu}h\right),
\end{equation}
with distance to the critical point $\delta r = \bar{r}_{0} - \bar{r}_{0,c}$ and
scaling parameter $b$. Note, $\delta r$ can in fact be realized
by either changing $\bar{r}_{0}$ or by changing $\bar{r}_{0,c}$ that depends
on $u.$ Thus, for the discussion of scaling behavior, it is sufficient
to include only two scaling fields $R=\left(\delta r,h\right)$. The
choice $b=\delta r^{-\nu}$ implies for $h=0$ that $\phi_{{\rm eq}}\left(\delta r\right)\propto\delta r^{\beta}$,
where $\beta$ is the order-parameter exponent. $\nu$ is the correlation
length exponent with equilibrium correlation length 
\begin{equation}
\xi \propto\delta r^{-\nu}.
\end{equation}
In analogy, right at the critical point and for finite field follows
with $b=h^{-\nu/\left(\beta\delta\right)}$ that $\phi_{{\rm eq}}\left(h\right)\propto h^{1/\delta}$.
Those are the well known scaling relations in equilibrium.

In a non-equilibrium setting the order parameter will, on the one
hand, depend on time, which transforms under scaling as $t\rightarrow b^{-z}t$. Here, $z$ is the dynamic scaling exponent that relates typical time
and length scales. We consider a regime where the dynamics is dominated
by the coupling to the bath. In this case, to leading order in $1/N$
we find that like in the equilibrium case discussed previously (see Eq.~\eqref{eq:15}) that 
\begin{equation} 
\label{eq:16}
 z=2/\alpha
\end{equation}
 is determined by the spectral function
$\eta\left(\omega\right).$ On the other hand, our quench protocol
implies that the order parameter depends on the initial distance $\delta r_{i}$ and
the final distance $\delta r_{f}$ to the critical point as well as on the
initial and final fields $h_{i}$ and $h_{f}$. It just holds that 
\begin{equation}
\phi\left(t,R_{i}\left(1\right),R_{f}\left(1\right)\right)=b^{-\beta/\nu}\phi\left(b^{-z}t,R_{i}\left(b\right),R_{f}\left(b\right)\right),\label{eq:scalin OP}
\end{equation}
where $R_{f}\left(b\right)=\left(b^{1/\nu}\delta r_{f},b^{\beta\delta/\nu}h_{f}\right)$
and $R_{i}\left(b\right)=\left(b^{\kappa/\nu}\delta r_{i},b^{\kappa\beta\delta/\nu}h_{i}\right)$.
The scaling dimension of the order parameter $\phi$ and of the final values
$\delta r_{f}$ and $h_{f}$ continue to take their equilibrium values
$\beta$, $\nu^{-1}$, and $\beta\delta/\nu$ respectively. This is
a consequence of the fact that the system approaches equilibrium after
very long time scales
\begin{equation}
\lim_{t\rightarrow\infty}\phi\left(t,R_{i},R_{f}\right)=\phi_{{\rm eq}}\left(R_{f}\right),
\end{equation}
independent on the initial values $\delta r_{i}$ and $h_{i}$. Thus,
the corresponding scaling dimension are the same as in equilibrium.
There is, however, no reason why the scaling dimension of $\delta r_{i}$
and $h_{i}$ should also be equal to the equilibrium values. This
is reflected in the new exponent $\kappa$ that enters Eq.~\eqref{eq:scalin OP}.
At this point it is not obvious why the same exponent $\kappa$ modifies
the scaling dimension of $\delta r_{i}$ and $h_{i}$. This will only
become clear in our explicit analysis of the theory. To demonstrate
this and to determine $\kappa$ is one of the goals of this paper.

Let us further analyze implications of this scaling behavior. For
simplicity, we first consider the case of a quench right to the critical point $\delta r_{f}=0$ in vanishing field $h_{i}=h_{f}=0$.
In this case Eq.~\eqref{eq:scalin OP} simplifies to 
\begin{equation}
\phi\left(t,\delta r_{i}\right)=b^{-\beta/\nu}\phi\left(b^{-z}t,b^{\kappa/\nu}\delta r_{i}\right).
\end{equation}
with $b^{-z}t=t_{\gamma}$ and $t_{\gamma}$ being a microscopic time
scale. We can therefore express the order parameter in terms of a scaling function as
\begin{equation}
\label{eq:17}
\phi\left(t,\delta r_{i}\right)=t^{-\frac{\beta}{\nu z}}\Psi(t/t^{*}),
\end{equation}
where $t^{*}\propto\delta r_{i}^{-\frac{\nu z}{\kappa}}$ and $t$, $t^*$ are in units of $t_\gamma$. For $t\gg t^{*}$
the scaling function is expected to approach a constant $\Psi(x\gg 1) \rightarrow \text{const.}$
and the order parameter decays adiabatically according to a power-law characterized by equilibrium exponents
\begin{equation}
\label{eq:18}
\phi\left(t\gg t^{*}\right)\propto t^{-\beta/\left(\nu z\right)}\,.
\end{equation}
On the other hand, for short times $t\ll t^{*}$ one expects that
$\phi\propto\delta r_{i}^{\beta}$ such that $\Psi(x\ll 1 )\propto x^{\kappa\beta/(\nu z)}$.
In this regime follows 
\begin{equation}
\phi\left(t\ll t^{*}\right)\propto t^{\left(\kappa-1\right)\beta/\left(\nu z\right)} \,.
\end{equation}
As shown below the exponent $\kappa$ varies with $z$ and is for
some $z$-values larger than unity, implying that the order parameter rises
for intermediate time scales, while for other $z$ values $\kappa<1$
such that a slower decay of $\phi$ governs the regime up to the crossover
scale $t^{*}$. Note that $t^*$ diverges for shallow quenches where $\delta r_{i}$
is small. An analogous behavior occurs in the regime where we perform
a quench of the field $h_{i}$ to $h_{f}=0$ for a system at the critical
point: $\delta r_{i}=\delta r_{f}=0$. The characteristic time scale
is now given as $t^{*}\propto h_{i}^{-\frac{\nu z}{\beta\delta\kappa}}$.
The resulting time dependence for $t\ll t^{*}$and $t\gg t^{*}$ are
the same as before. 

Let us now turn to the correlation and response functions. In equilibrium,
the retarded and Keldysh Green's functions only depend on the difference
$t-t'$ of the two time variables and obey at the critical point the
established scaling behavior 
\begin{equation}
G_{\text{eq}}^{R(K)}\left(q,t,t'\right)=\frac{F_{{\rm eq}}^{R(K)}\left(q^{z}\left(t-t'\right)/\gamma^{z/2}\right)}{q^{2-z-\eta}\gamma^{z/2}},\label{eq:GRscaling}
\end{equation}
with scaling functions $F_{{\rm eq}}^{R}\left(x\right)$ and $F_{{\rm eq}}^{K}\left(x\right)$.
The factor $\gamma^{z/2}$ occurs to render the scaling functions
and their arguments dimensionless. As discussed, this behavior is
only valid for time scales sufficiently long to ensure that the dynamics
is dominated by the bath and it is justified to neglect $\omega^2$ compared to $|\omega|^\alpha$ in the bare propagators. 

Considering now the out-of-equilibrium dynamics after the quench to
the critical point, correlations and response depend in general on
both time variables. This gives rise to the additional dimensionless
ratio $t/t'$, compared to scaling in equilibrium. Thus, we expect
from dimensional considerations the following behavior: 
\begin{equation}
G^{R(K)}\left(k,t,t'\right)=\left(\frac{t}{t'}\right)^{\theta(\theta')}\frac{F^{R(K)}\left(q^{z}t/\gamma^{z/2},t/t'\right)}{q^{2-z}\gamma^{z/2}}.\label{eq:scaling}
\end{equation}
Here, the scaling functions $F^{R}$ and $F^{K}$ are chosen such
that they depend only weakly on the ratio $t'/t$ if $t\gg t'$. The
possibility of a singular dependence on $t/t'$ is captured by the
exponents $\theta$ and $\theta'$, respectively. Below we will determine
scaling laws that relate the exponents $\theta$, $\theta'$, and $\kappa$
in such a way that only one new independent exponent emerges. Thus,
the post quench dynamics is governed by a single critical exponent that cannot be
expressed in terms of equilibrium exponents.

\section{Quench from the disordered phase}
\label{sec:quench-from-disord}
In what follows we present the solution of the coupled large-$N$ equations (see Eqs.~\eqref{eq:19}-\eqref{eq:GKrenorm}) for the post-quench non-equilibrium dynamics at the quantum critical point starting the quench from the disordered phase. The quench protocol is indicated in by $A \rightarrow C$ in Fig.~\ref{fig:2}.
The effective time dependent mass in Eq.~\eqref{eq:largeN-time-dep-mass}
simplifies to
\begin{equation}
r(t)=\bar{r}_{0,f}+\frac{u}{2}\int\frac{d^{d}q}{\left(2\pi\right)^{d}}iG_r^{K}(q,t,t).
\end{equation}
In case of a quench to the critical point, $\bar r_{0,f}$ takes the value
$\bar r_{0,c}$ in Eq.~\eqref{eq:r0equil}. It is convenient to express $\bar r_{0,c}$ in terms of the
equilibrium Keldysh function such that 
\begin{equation}
r\left(t\right)=\frac{u}{2}\int\frac{d^{d}q}{\left(2\pi\right)^{d}}\left(iG_r^{K}\left(q,t,t\right)-iG_{{\rm eq}}^{K}\left(q\right)\right)\,.\label{eq:selfconst}
\end{equation}
For times larger than the microscopic time scales  
the Keldysh function obeys scaling. We refer to Appendix~\ref{sec:short-time-scales} for a discussion of the various short-time scales in the problem. Based on dimensional arguments, we make the following generalized\emph{ light-cone} ansatz
\begin{equation}
r\left(t\right)=\frac{\gamma a}{t^{2/z}},\label{eq:lcone}
\end{equation}
valid for $t$ larger than the microscopic time scale. The prefactor $\gamma$, that determines the strength of the coupling to the bath,
was chosen to ensure that the light-cone amplitude $a$ is dimensionless. This ansatz corresponds to a time-dependent correlation length
\begin{equation}
\xi\left(t\right)=\left|r\left(t\right)\right|^{-2}=\frac{1}{(\gamma\left|a\right|)^{1/2}}t^{1/z},
\end{equation}
reflecting the fact that after time $t$ perturbations propagate a distance $\xi\left(t\right)$. For ballistic propagation with $z=1$ this corresponds to the usual light-cone propagation during time $t$. In general, $\xi(t)$ is the length scale on which the signal of a perturbation can propagate. Among others, this implies that the system will eventually reach a state with diverging correlation length. This is expected as the coupling to the heat bath ensures that the system equilibrates at the quantum critical point since the bath temperature is zero.

We now demonstrate that Eq.~\eqref{eq:lcone} indeed leads to a self-consistent solution of the coupled large-$N$ equations and determine the light-cone amplitude $a$. Let us next explore some consequences of such a time dependent mass $r(t)$.

\subsection{Light-cone amplitude $a$ and universal short-time exponent $\theta$}
\label{sec:light-cone-amplitude}
We first establish a relation between the light-cone amplitude $a$
of Eq.~\eqref{eq:lcone} and the exponents $\theta$ and $\theta'$ of the retarded and Keldysh Green's
functions in Eq.~\eqref{eq:scaling}. We will show that $\theta = \theta'$, which is in contrast to the regime near classical phase transitions~\cite{Janssen-ZPhysB-1989,1742-5468-2012-01-P01014} and to the dynamics in an isolated, quantum system~\cite{PhysRevB.91.220302}. This relation can easily be determined from an analysis
of the Dyson equation of $G^R$ in the intermediate regime $t_{\gamma}\ll t\ll t_q \equiv \left(\sqrt{\gamma}/q\right)^{z}$
and $t\gg t'$. The leading corrections that follow from Eqs.~\eqref{eq:DysonR} and~\eqref{eq:DysonK} in this limit are given by
\begin{eqnarray}
\label{eq:8}
\delta G^{R}_r\left(k, t,t^{\prime}\right) & \approx & \int_{t_{\gamma}}^{t}ds g^{R}\left(k, t-s\right)\frac{\gamma a }{s^{2/z}} g^{R}\left(k,s-t^{\prime}\right) \\
\delta G^{K}_r\left(k, t,t'\right) & \approx & \int_{t_{\gamma}}^{t}ds g^{R}\left(k,t-s\right) \frac{\gamma a }{s^{2/z}} g^K\left(k,s,t'\right)\,.\label{eq:dGlead}
\end{eqnarray}
Here, the lower integratl limit $t_{\gamma}$ appears since the power-law decay of $r\left(t\right)\propto t^{-2/z}$ sets in only beyond this
time scale.

We first consider the correction to the retarded function. For frequencies $\omega\gg\left(q/\sqrt{\gamma}\right)^{z}$, which correspond to times $t\ll\left(\sqrt{\gamma}/q\right)^{z}$, the bare retarded propagator $g^R$ at the critical point is local in space and entirely governed by the bath
\begin{equation}
g^{R}\left(q,\omega\right)\approx\frac{1}{\delta\eta\left(\omega\right)}\,.
\label{eq:GR0short}
\end{equation}
Here, $\delta\eta\left(\omega\right)$ is the frequency dependent contribution
of the bath function $\eta\left(\omega\right)$ (see Eq.~\eqref{eq:deta}).
Performing the inverse Laplace transformation we obtain 
\begin{eqnarray}
g^{R}\left(q,t\right) & \approx & -\frac{\sin(\pi/z)}{\gamma \Gamma(2/z)} t^{2/z-1}.
\end{eqnarray}
If we insert this into Eq.~\eqref{eq:dGlead}, it follows for $t' > t_\gamma$ that 
\begin{eqnarray*}
\delta G^{R}_r\left(t,t'\right) & = & \frac{a\sin^{2}(\pi/z)}{\gamma\Gamma(2/z)^{2}} \int_{t'}^{t}ds\frac{\left(t-s\right)^{2/z-1}\left(s-t'\right)^{2/z-1}}{s^{2/z}}.
\end{eqnarray*}
The integral can be performed analytically. Evaluating it in the limit
$t\gg t'$ yields 
\begin{equation}\label{eq:delta Gr}
\delta G^{R}_r\left(t,t'\right)=\frac{a\sin^{2}(\pi/z)}{\gamma\Gamma(2/z)^{2}}t^{2/z-1}\log\frac{t}{t'} \,.
\end{equation}
To leading order it follows for the retarded propagator
\begin{eqnarray}
\label{eq:21}
G^{R}_r\left(t,t'\right) & = & g^{R}\left(t-t'\right)\Bigl(1+\theta\log\frac{t}{t'}+ \ldots \Bigr) \,,
\end{eqnarray}
with 
\begin{equation}
\theta=-\frac{a\sin(\pi/z)}{\Gamma(2/z)}\,.
\label{eq:exponent relation}
\end{equation}
This result demonstrates that a time-dependent mass $r(t)$ leads to aging behavior of the retarded Green's function, i.e. $G_r^R(t,t')$ depends on both time arguments $t$ and $t'$ separately. 
If we include higher order corrections, an analysis along the same
lines yields higher powers of the logarithm with appropriate coefficients
allowing us to exponentiate the logarithm if $a$ is small. We can
then put $G^{R}_r\left(t,t'\right)$ in the scaling form of Eq.~\eqref{eq:scaling}
and find that $\theta$ is indeed the exponent that appears in the scaling form.
Thus, the dimensionless amplitude $a$ of the time-dependent mass determines the
exponent $\theta$ of the retarded Green's function. This analysis
was performed under the assumption of negligible $q-$dependence of
the retarded propagator. Including this $q-$dependence is technically
slightly more involved, but leads to the same behavior since the $\log(t/t')$-time
dependence is still the dominant one in the limit $t\gg t'$.

Let us now discuss the interaction correction to the short-time behavior of the Keldysh function using Eq.~\eqref{eq:dGlead}. We need to evaluate the short time behavior of the non-interacting Keldysh function $g^{K}(t,t')$, for which it is useful to introduce a double Laplace-transformation (see Ref.~\onlinecite{1742-5468-2012-01-P01014})
\begin{equation}
 g^{K}(q,\omega,\omega')= \int_0^\infty dt dt' e^{i(\omega+i0^+)t}e^{i(\omega'+i0^+)t'}g^{K}(q,t,t') \,.
\end{equation}
As we show in Appendix~\ref{app:post-quench G} the Keldysh function can be expressed in terms of the memory function $M(q, \omega, \omega')$ as $g^K(q,\omega, \omega') = M(q, \omega, \omega') g^R(q, \omega) g^R(q,\omega')$. Using the deep quench result for the memory funcion in Eq.~\eqref{eq:memodq} at $T=0$, this takes the form
\begin{align}
\label{eq:24}
g^{K}(q,\omega,\omega') &=  i\frac{{\rm sign}(\omega) \delta\eta(\omega)+{\rm sign}(\omega') \delta\eta(\omega')}{\omega+\omega'+i0^{+}} \nonumber \\
 & \quad \times g^{R}(q,\omega)g^{R}(q,\omega') \,.
\end{align}
The limit $t\gg t'$ corresponds to $\omega'\gg\omega$.
In this limit $\delta\eta(\omega')$ is always large compared to $\delta\eta(\omega)$
and it follows 
\begin{equation}
g^{K}(q,\omega,\omega'\gg\omega)\simeq i\frac{\delta\eta(\omega')}{\left|\omega'\right|}g^{R}(q,\omega)g^{R}(q,\omega').
\end{equation}
Using Eq.~\eqref{eq:GR0short} for the retarded function in the short time limit, it immediately follows 
\begin{equation}
g^{K}(q,\omega,\omega'\gg\omega)\simeq\frac{i}{|\omega'|}g^{R}(q,\omega) \,.
\end{equation}
The back transformation is now straightforward and, using that $1/|\omega'|$ turns into a constant upon Laplace transformation, we find
\begin{equation}
g^{K}(q,t\gg t')\simeq g^{R}(q,t)=-\frac{\sin(\pi/z)}{\gamma\Gamma(2/z)}t^{2/z-1}.\label{eq:Keldysh-short-times}
\end{equation}
In the short time limit and for $t\gg t'$, the Dyson equations for the retarded and
Keldysh functions thus only differ in the bare values of the two functions. Since we just demonstrated that $g^{K}(q,t\gg t')$ and $g^{R}(q,t)$
have the same $t-$dependence in the relevant regime, both Dyson equations have the same solutions. This immediately yields $\theta'=\theta$
for the exponent $\theta'$ of the Keldysh function introduced in Eq.~\eqref{eq:scaling}.

\subsection{Long-time behavior of $G^{K}(q,t,t)$}
\label{sc:long-time-GK}
We next determine the long time decay of the Green's functions $G^R$ and $G^K$ of a system
with time-dependent effective mass $r\left(t\right)$ as given in Eq.~\eqref{eq:lcone}. To this end we analyze the equation of motion~\eqref{eq:eomlN}, in the long time limit. From the equation of motion follows
\begin{eqnarray}
\left(\partial_{t}^{2}+r\left(t\right)+q^{2}\right)\boldsymbol{\varphi}\left(q,t\right) & = & \int_{-\infty}^{t}ds\delta\eta(t-s)\boldsymbol{\varphi}(q,s).\nonumber\\
& &\label{eq:eomlN-1}
\end{eqnarray}
In order to analyze the impact of the time dependence of the correlation
length, we perform an expansion for small $r\left(t\right)$. For
the specific ansatz in Eq.~\eqref{eq:lcone}, this amounts to an expansion
in the dimensionless light-cone amplitude $a$: 
\begin{equation}
\boldsymbol{\varphi}(q,t)=\boldsymbol{\varphi}_{{\rm eq}}(q,t)+\boldsymbol{\varphi}_{1}(q,t)+\mathcal{O}(a^{2}),\label{eq:long-time-expansion-varphi}
\end{equation}
where $\boldsymbol{\varphi}_{{\rm eq}}(q,t)$ obeys the equation of motion for $r\left(t\right)=0$, i.e. the equation of motion in equilibrium
at the critical point. We can split the integral on the right hand
side of Eq.~\eqref{eq:eomlN-1} into 
\begin{align}
\boldsymbol{\Xi}\left(t\right)=\int_{-\infty}^{0}ds\delta\eta(t-s)\boldsymbol{\varphi}(q,s)
\end{align}
and a contribution that only contains the post-quench dynamics of
$\boldsymbol{\varphi}\left(q,t\right).$ As far as the post-quench
dynamics is concerned, $\boldsymbol{\Xi}\left(t\right)$ acts as an inhomogeneity
in Eq.~\eqref{eq:eomlN-1}. At long times we can neglect this inhomogeneity
since $\delta\eta\left(t\rightarrow\infty\right)$ vanishes sufficiently
rapidly. This determines the leading perturbative correction to
\begin{equation}
\boldsymbol{\varphi}_{1}(q,t)=-\int_{t_{0}}^{t}dsG_{\text{eq}}^{R}(t-s)r(s)\boldsymbol{\varphi}_{{\rm eq}}(q,s)\,,
\end{equation} 
where $G^R_{\text{eq}}(t-t') = - i \theta(t-t')\langle[\varphi_{\text{eq}}(t), \varphi_{\text{eq}}(t')]\rangle$ (suppressing the field index $l$ again). 

Using the definition of $G^{R/K}$ in Eqs.~\eqref{eq:defGR} and~\eqref{eq:defGK} we can now determine how the Green's functions relax towards their equilibrium expressions. Let us first consider the equal-time Keldysh function $G_r^K(q,t,t)$, which turns out to be of particular importance for the self-consistent solution of the large-$N$ equation. Using Eq.~\eqref{eq:long-time-expansion-varphi} leads to
\begin{equation}
G_r^{K}(q,t,t)=G_{{\rm eq}}^{K}(q)+\delta G_r^{K}\left(q,t,t\right),
\end{equation}
where 
\begin{equation}
\delta G_r^{K}\left(q,t,t\right)=-2\int_{t_{\gamma}}^{t}dsG_{\text{eq}}^{R}(t-s)r(s)G_{{\rm eq}}^{K}(t-s).
\end{equation}
To evaluate this expression we express the retarded function in Laplace
and the equilibrium Keldysh function in Fourier space and obtain 
\begin{align}
\delta G_r^{K}\left(q,t,t\right) & =  -2i\int\frac{d\omega d\omega'}{2\pi^{2}}{\rm Im} \bigl[ G_{\text{eq}}^{R}(\omega) \bigr] G_{\text{eq}}^{K}(\omega')\nonumber \\
 & \quad \times \int_{t_{\gamma}}^{t}ds \, r(s)e^{-i(\omega-\omega')(t-s)}\,.
\end{align}
We want to evaluate this integral in the limit of large $t$. The typical frequency scale of the Green's function is $\omega_q \approx t_q^{-1} = q^{z}/\gamma^{z/2}$. For times $t\gg \omega_q^{-1}$ the integrand is in general highly
oscillatory, except for contributions that stem from the upper limit of the integration over time. This yields up to leading order in $t$ 
\begin{equation}
\int_{t_{\gamma}}^{t}dse^{-i\left(\omega-\omega'\right)\left(t-s\right)}r\left(s\right)\approx\frac{ir(t)}{\omega-\omega'} \,.
\label{eq:long-time-GK-approximation}
\end{equation}
Performing the remaining frequency integration leads to 
\begin{equation}
G_r^{K}(q,t,t)=G_{{\rm eq}}^{K}(q)+\frac{2ir\left(t\right)}{c_{K}q^{4-z}\gamma^{z/2}}\label{eq:GKlongtime}
\end{equation}
with numerical coefficient 
\begin{align}
\label{eq:9}
 c_{K}=\frac{4\sin(\pi z/2)}{z\left(2-z\right)\sin^{z/2}(\pi/z)}.
\end{align}
This result demonstrates that the equal-time correlation function decays for large times according to a power law. Including higher order corrections in the Dyson equation, leads to terms of order $r(t)^2$. Thus, the term $G_1^K(q,t,t,)= 2ir(t)/\bigl( c_{K}q^{4-z}\gamma^{z/2}\bigr)$ appearing in Eq.~\eqref{eq:GKlongtime} is the slowest decaying correction at large times.  This is an interaction effect, because the bare correlation function $g^{K}(q,t,t)$ decays for finite momenta always exponentially $\propto\exp\left(-q^{z}t/\gamma^{z/2}\right)$ to its value in equilibrium. Critical fluctuations lead to a significant slowing down of the equilibration of the system.

\subsection{Self-consistent determination of the light-cone amplitude}
\label{sec:self-consistent-solution}
In this section, we demonstrate that the ansatz for the time-dependent mass $r(t)$ in Eq.~\eqref{eq:lcone} indeed leads to a self-consistent solution of the coupled large-$N$ equations and we show how to determine $a$. Let us begin with some general remarks about Eq.~\eqref{eq:selfconst} that we want to solve self-consistently 
\begin{equation}
\frac{a\gamma}{t^{2/z}}=\frac{uK_{d}}{2}\int_{0}^{\varLambda}dq\, q^{3-z-\epsilon}\left(iG_r^{K}(q,t,t)-iG_{{\rm eq}}^{K}(q)\right)\,.
\label{eq:explicit-self-consistent}
\end{equation}
Here, $K_{d}=\Gamma(d/2)/[2\pi^{d/2}(2\pi)^{d}]$ takes into account the integration over angles and 
\begin{equation}
\epsilon = 4-d - z \,.
\end{equation}
We want to determine the dimensionless light cone amplitude $a$
self-consistently for small $\epsilon$. Since $u$ will turn out
be also of order $\epsilon$ (see Eq.~\eqref{eq:fixpoint-u} below) it
is tempting to speculate that one only has to expand $G_r^{K}$ in $a$.
To leading order it would then suffice to insert the bare Keldysh
function $g^{K}(q,t,t)$ in Eq.~\eqref{eq:explicit-self-consistent}.
Since $g^{K}$ decays exponentially to $G_{{\rm eq}}^{K}$ the
integral is convergent in the limit $\varLambda\rightarrow\infty$.
However, as was shown in the previous paragraph, first order corrections
to the Keldysh function decay much more slowly. Using Eq.~\eqref{eq:GKlongtime} such corrections $G_{1}^{K}(q,t,t) \propto r(t)/q^{4-z}$
are proportional to $q^{-4+z}$. Upon integration this generates terms that behave as $1/\epsilon$ for small $\epsilon$:
\begin{equation}
\int_{q_{0}}^{\varLambda}dq\frac{1}{q^{1+\epsilon}}=-\frac{1}{\epsilon}\left(\varLambda^{-\epsilon}-q_{0}^{-\epsilon}\right),\label{eq:1over-eps-term}
\end{equation}
where $q_{0}$ is some appropriate lower cut-off that we elaborate on below. Such a term, multiplied
with $r\left(t\right)\propto a$ will be of same order $\mathcal{O}(\epsilon^0)$ as the bare $g^{K}$. Thus, in the expansion of $G_r^{K}$ in $a$ we have to keep those slowly decaying terms proportional to $q^{-4+z}$. From $G_{1}^{K}$ in Eq.~\eqref{eq:GKlongtime} follows that large times $t$ correspond to large momenta $q \gg q_0 \equiv \gamma^{1/2}/t^{1/z}$, which naturally introduces the lower cut-off $q_0$ in Eq.~\eqref{eq:1over-eps-term}. As we have previously shown in Sec.~\ref{sec:light-cone-amplitude}, in the opposite limit of small momenta $q \ll q_0 \gamma^{1/2}/t^{1/z}$ the Keldysh-function becomes momentum independent (see Eq.~\eqref{eq:Keldysh-short-times}) and cannot generate contributions that behave as $1/\epsilon$.

\begin{figure}[tb]
  \centering
  \includegraphics[width=\linewidth]{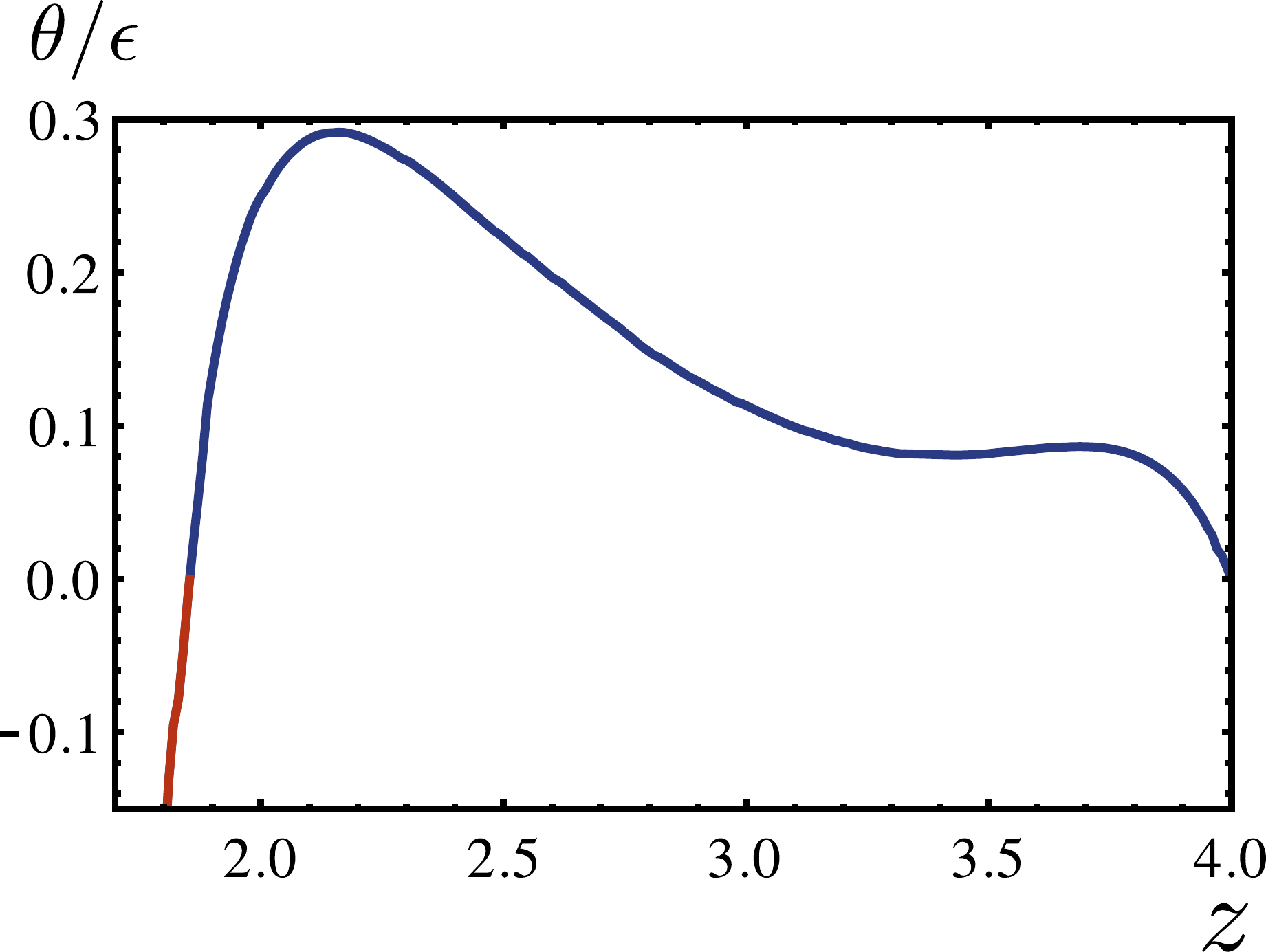}
  \caption{(Color online) Prethermalization exponent $\theta$ as a function of dynamic critical exponents $z$. Here, $\epsilon = 4 - d - z$ and $z = 2/\alpha$ is determined by the form of the bath spectral function at low energies $\text{Im} \eta(\omega) \propto |\omega|^{\alpha}$. For $\theta < 0$ one observes an underdamped approach of the free Keldysh Green's function $G_0^K(t,t)$ towards equilibrium. The transition from overdamped to underdamped behavior occurs close the location of the sign change.    }
  \label{fig:4}
\end{figure}

With these considerations, we are able to solve the large-$N$ equations self-consistently. We expand $G_r^{K}=g^{K} + G_{1}^{K}$ in small $r(t)$ to find
\begin{eqnarray}
r(t) & = & \frac{uK_{d}}{2}\int_{0}^{\infty}dq\, q^{3-z-\epsilon}\left(ig^{K}(q,t,t)-iG_{{\rm eq}}^{K}(q)\right)\nonumber \\
 & + & \frac{uK_{d}r(t)}{c_{K}\epsilon\gamma^{z/2}}\Bigl(\varLambda^{-\epsilon}-\Bigl(\frac{\gamma^{1/2}}{t^{1/z}}\Bigr)^{-\epsilon}\Bigr).
\end{eqnarray}
In the first integral we already took the limit $\varLambda\rightarrow\infty$ because $g^K$ approaches $G^K_{\text{eq}}$ exponentially quickly and the integral is thus convergent at the upper limit. To proceed, we use the scaling form of Eq.~\eqref{eq:GRscaling} and
introduce a dimensionless integration variable $x=q^{z}t/\gamma^{z/2}$ to obtain 
\begin{equation}
\label{eq:25}
\frac{a\gamma}{t^{2/z}}=\frac{uK_{d}}{2z\gamma^{z/2}}\frac{\gamma C_{0} t^{\epsilon/z}}{\gamma^{\epsilon/z} t^{2/z}}+\frac{ua\gamma K_{d}}{c_{K}\epsilon\gamma^{z/2}t^{2/z}}\Bigl(\varLambda^{-\epsilon}-\frac{t^{\epsilon/z}}{\gamma^{\epsilon/2}}\Bigr)\,,
\end{equation}
where we have introduced the dimensionless integral 
\begin{equation}
C_{0}=\int_{0}^{\infty}dx\, x^{2/z-1}\left(if^{K}(x,1)-iF_{{\rm eq}}^{K}\right)\,.
\label{eq:C0}
\end{equation}
The equation~\eqref{eq:25} must hold for all times $t \gg t_{\gamma}$, and therefore terms with and without $t^{\epsilon/z}$ must cancel separately. This allows to extract the two conditions
\begin{eqnarray}
u & = & u^{*} \equiv \frac{c_{K}\gamma^{z/2}\varLambda^{\epsilon}}{K_{d}}\epsilon,\label{eq:fixpoint-u}\\
a & = & \frac{c_{K}C_{0}}{2z}\epsilon.\label{eq:condition-for-a}
\end{eqnarray}
The first equation determines the deep-quench fixed point value of $u$. It does not imply that $u$ must be tuned precisely to this value. Instead, it should be understood as a consequence of using the deep-quench scaling form of $G_r^{K}$. Non-universal contributions in $G_r^{K}$ will renormalize $u$, which is fully consistent with our renormalization group based reasoning in Ref.~\onlinecite{PhysRevLett.113.220401}. The second equation~\eqref{eq:condition-for-a} determines the value of $a$ and, using Eq.~\eqref{eq:exponent relation}, the value of the new critical exponent 
\begin{equation}
\label{eq:22}
\theta=- \frac{\sin(\pi/z) a}{\Gamma(2/z)} = -\frac{c_{K}\sin(\pi/z)}{2z\Gamma(2/z)}C_{0}\epsilon.
\end{equation}
Note, that exactly the same value was obtained in our previous renormalization-group analysis in the limit $N \rightarrow \infty$~\cite{PhysRevLett.113.220401}, where we have shown that the integral $C_0$ can be performed analytically in the case of an ohmic bath. For general dynamic critical exponents $1<z<4$ we evaluate it numerically and the result is presented in Fig.~\ref{fig:4}.

 The relation between the light-cone amplitude $a$ and the retarded short-time critical exponent $\theta$ is identical for a classical and a quantum system, however, the $z$-dependence of $\theta$ is completely different due to the fact that quantum fluctuations are only present in the quantum system. Quantum fluctuations influence $\theta$ in two ways: (i) via the numerical constant $c_K$ in Eq.~\eqref{eq:9} and (ii) via the result of the dimensionless integral $C_0$ over the Keldysh functions in Eq.~\eqref{eq:C0}. 

The constant $c_K$ determines the fixed point $u^*$ as well as the long-time decay of $G_r^K$. For a quantum system, $c_K$ is positive for $z\leq 4$ and linearly approaches zero for $z\rightarrow 4$. The constant $C_0$ is given by the difference of the post quench Keldysh scaling function $f^K(x)$ and the equilibrium Keldysh scaling function $F^K_{\text{eq}}$. For a classical system, the system is always overdamped and $f^K$ is smaller than $F^K_{\text{eq}}$ for all values of $z$. Hence, the exponent for a classical system is always positive $\theta_{cl}> 0$. For an ohmic bath, for example, one finds $\theta_{cl}=\epsilon_{cl}/4$with $\epsilon_{cl} = 4 - d$~\cite{Janssen-ZPhysB-1989}. For a quantum system, however, $f^K$ can exhibit oscillations, in particular for super-Ohmic bath spectral functions where $z \lesssim 2$ the Keldysh function becomes underdamped. This leads to a sign change of $\theta$ at $z\approx 1.8 $. 

While the classical result of the exponent for an ohmic bath is $\theta_{cl}=\epsilon_{cl}/4$, which is the same value as in the quantum case for $z=2$: $\theta(z=2) = \epsilon/4$, this is only a coincidence, since the constants $c_K$ and $C_0$ take different values for $T=0$ and $T\rightarrow \infty$. For an ohmic bath, one can therefore still distinguish the quantum post-quench dynamics from the classical one by analyzing, for example, the long-time behavior of the Keldysh-function. Moreover, since $\epsilon$ depends on $z$, we find the same short time exponent for classical and quantum systems in different dimensions. For example, we predict $\theta_{cl} = 1/4$ for a classical system in contact with an Ohmic bath in $d=3$ dimensions, while $\theta = 1/4$ for a quantum system in contact with an Ohmic bath in $d=1$ dimension. 

\subsection{Distribution function $n(t,\omega)$}
\label{sec:distribution-fct}
The analysis of the long-time behavior of the equal-time Keldysh function $G^K(q,t,t)$ in Sec.~\ref{sc:long-time-GK} can be straightforwardly extended to different times $t,t'$ as well as to retarded function $G^R(q,t,t') $. We show in detail in Appendix~\ref{app:long-time-limit} that in the limit where both time arguments are large compared to the typical mode time $t_q = q^{-z}\gamma^{z/2}$, but the relative time is small $t-t'\ll t_q$, this leads to
\begin{align}
 G_r^K(q,t,t')=& G^K_{\text{eq}}(q,t-t')-2r\left(\frac{t+t'}2\right) C^K(q,t-t') \label{eq:long-time-GK-tt'}\\
 G_r^R(q,t,t')=& G^R_{\text{eq}}(q,t-t') \label{eq:long-time-GR-tt'} \\
 &-4i \theta(t-t')r\left(\frac{t+t'}2\right) C^R(q,t-t') \nonumber
\end{align}
where we have defined the functions
\begin{align}
 C^K(q,t)= &\int \frac{d\omega}{2\pi}{\rm Re} \bigl[ G^R_{\text{eq}}(q,\omega) \bigr] G^K_{\text{eq}}(q,\omega)e^{-i\omega t} \\
 C^R(q,t)=&\int \frac{d\omega}{2\pi}{\rm Re} \bigl[ G^R_{\text{eq}}(q,\omega) \bigr] {\rm Im}\bigl[ G^R_{\text{eq}}(q,\omega) \bigr] e^{-i \omega t}.
\end{align}
By performing a Wigner-transformation of Eqs.~\eqref{eq:long-time-GK-tt'} and~\eqref{eq:long-time-GR-tt'} and by using the fluctuation-dissipation theorem for the equilibrium Keldysh function $G^K_{\text{eq}} = 2 i \coth[\omega/(2 T)] \text{Im} G^R_{\text{eq}}$, one can express $G_r^K$ after a few steps as 
\begin{align}
\label{eq:11}
 G_r^K(q,t_a, \omega)= 2i& \coth\left(\frac{\omega}{2T} \right) \left[ 1+2r(t_{a})\rm{Re} G^R_{\text{eq}}(q,\omega)\right]\nonumber\\
&\times {\rm Im}G_r^R( q, t_{a},\omega)\,,
\end{align}
where $t_a = (t + t')/2$. This result shows, that we are in the limit of adiabatic relaxation, since $G_r^K$ can be written as
\begin{align}
 G_r^K(q,t_a, \omega)=&\coth\left(\frac{\omega}{2T} \right)\nonumber\\
 &\times \frac{2i{\rm Im}\eta(\omega)}{[ q^2+r(t_a)+{\rm Re}\eta(\omega)]^2+[{\rm Im}\eta(\omega)]^2}.
\end{align}
Further, the similarity between $G^R_r$ and $G^K_r$ suggest to connect them via the fluctuation-dissipation theorem and to introduce a distribution function $n(t_a,\omega)$ via
\begin{equation}
 G_r^K(q,t_a, \omega)=  2 i \left[2n(t_a,\omega)+1\right] {\rm Im}G_r^R( q, t_{a},\omega).
\end{equation}
With $n(t_a,\omega)=n_{B}(\omega)+\delta n(t_a, \omega)$, where $n_{B}(\omega) $ is the Bose-distribution function, this yields for the correction 
\begin{equation}
\label{eq:5} 
\delta n(t_a,\omega)=\coth\left(\frac{\omega}{2T} \right) r(t_{a})\text{Re} \, G^R_{\text{eq}}(q,\omega).
\end{equation}
This expression is only valid for large frequencies, which corresponds to small relative times $|t-t'| \ll t_q$. For $t_a\rightarrow\infty$ the system relaxes to the thermal equilibrium state. Aging effects, however, lead to a significant slowing down of thermalization. Explicitly, one finds
\begin{align}
  \label{eq:32}
  \delta n &= \coth \Bigl( \frac{\omega}{2 T} \Bigr) \frac{\theta \Gamma(2/z)}{(|\omega| t_a)^{2/z}} \Bigl( \cos \frac{\pi}{z} + \frac{q^2}{\gamma |\omega|^{2/z}} \sin \frac{\pi}{z}  \Bigr) \,,
\end{align}
where both $q^2 |\omega|^{-2/z}/\gamma = (|\omega| t_q)^{-2/z} \ll 1$ as well as $(|\omega| t_a)^{-2/z} \ll 1$, since $t_a \gg t_q \gg |t-t'|$.

The correction to the Bose distribution is proportional to the short-time exponent $\theta$ and exhibits a (dominant) algebraic frequency dependence $\delta n \propto |\omega|^{-2/z}$ for $z \neq 2$ and $\delta n \propto |\omega|^{-2}$ in the ohmic case. This slow decay at large frequencies clearly shows that the system is not thermal and cannot be characterized by a temperature as the decay would then be exponential in $\omega$. The sign of $\delta n$ is positive for a (sub)-Ohmic bath $z \geq 2$, showing that even long after the quench there are an increased number of excitations at large frequencies present in the system as compared to equilibrium. For super-Ohmic bath spectral densities, however, the sign of $\delta n$ can change depending on the product of $\theta$, which becomes negative for $z \lesssim 1.8$, and the last term in Eq.~\eqref{eq:32}. This term stems from $\text{Re}\, G^R_{\text{eq}}$ and is positive (negative) for $z \geq 2$ ($z < 2$), except slightly below $z=2$ where the sign depends on the interplay of the $\cos(\pi/z) < 0$ and $(|\omega| t_q)^{-2/z} \sin(\pi/z) > 0$ with $(|\omega| t_q)^{-2/z} \ll 1$. More importantly, a negative sign of $\delta n$ implies that the density matrix of the system is not diagonal in the energy basis. The off-diagonal terms describe the presence of quantum coherence and thus make a straightforward interpretation of $\delta n$ in terms of a distribution function impossible. 

\begin{figure}[t!]
  \centering
  \includegraphics[width=\linewidth]{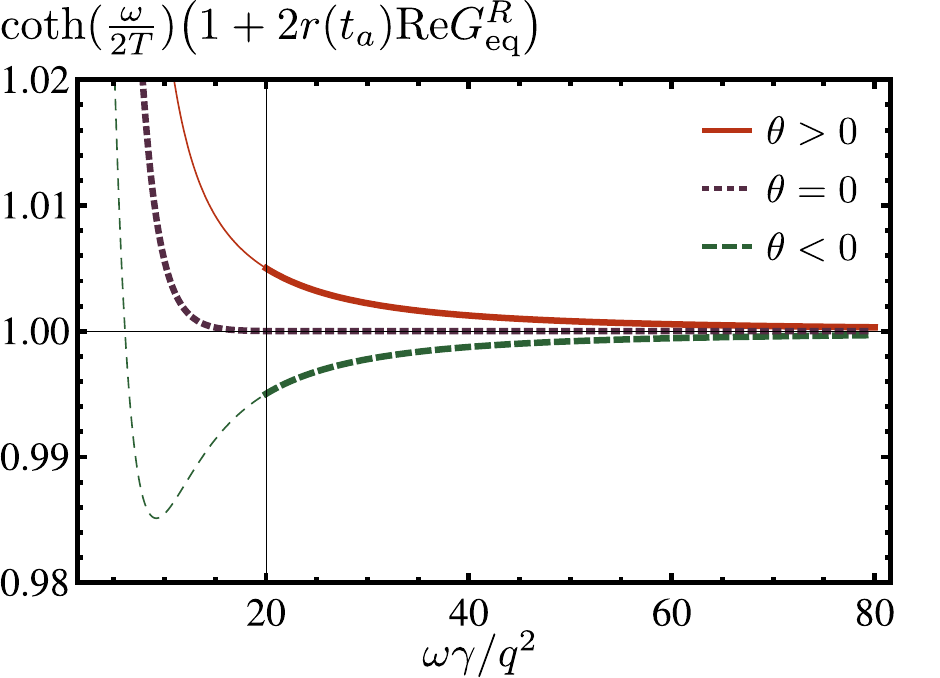}
  \caption{Visualization of the correction to the distribution function $\delta n(t_a, \omega)$ as a function of dimensionless variable $\omega \gamma/q^2$ (see Eq.~\eqref{eq:5}). Plot shows $2 n(t_a, \omega) + 1 = \coth(\frac{\omega}{2 T}) \bigl( 1 + 2 r(t_a) \text{Re} G^R_{\text{eq}} \bigr)$ (see Eq.~\eqref{eq:11}) as a function of the dimensionless variable $x = \gamma \omega/q^2$ for $z =2$, $\gamma T/q^2 = 1.7$ and $|\theta| \gamma/(q^2 t_a^{2/z}) = 1$. }
  \label{fig:5}
\end{figure}

In Fig.~\ref{fig:5} we illustrate this result for $2 n(t_a, \omega) + 1$ in the ohmic case, both for $\theta > 0$ as well as for $\theta < 0$ (although $\theta = \epsilon/4 > 0$ for $z=2$). We include $\theta< 0$ in the plot to illustrate the behavior in cases where $\theta < 0$ which is qualitatively the same. Above the upper critical dimension, one finds $\theta = 0$ and thus $\delta n = 0$. 

\subsection{Upper critical dimension}
\label{sec:upper-crit-dimens}
From the bare Green's function $g^{R}$ in Eq.~\eqref{eq:Goretarded} it follows immediately that the mean-field value of the exponent $\theta$ is zero. Therefore, $\theta$ vanishes above the upper critical dimension $d_{uc} = 4 - z$. In this section, we investigate the post-quench behavior if the system is exactly at the upper critical dimension $d = d_{uc}$. We can straightforwardly extend our previous result to this case. The large-$N$ equation for the mass reads 
\begin{equation}\label{eq.1/N-for-eps=0}
 r_{uc}(t)= \frac{uK_d}{2}\int dq q^{3-z}\left[ iG_{r_{uc}}^K(q,t,t)-iG^K_{\text{eq}}(q) \right],
\end{equation}
where we used that $d=d_{uc}$. In a next step we expand $G_{r_{uc}}^K$ in $r_{uc}(t)$. The important slowly decaying correction comes from terms going with $G^K_1(q,t,t)\propto q^{-4+z}$, generating logarithmic divergences after the $q$-integration. In Sec.~\ref{sc:long-time-GK} we calculated this correction, without making any assumption about the dimensionality of the system or the concrete decay of the time-dependent mass $r(t)$. Therefore, one can directly use the result of Eq.~\eqref{eq:GKlongtime}. Again, the condition $t> t_q = \gamma^{z/2}q^{-z}$ translates into a lower cut-off in the $q$-integral. For $q< q_0=\gamma^{1/2}/t^{z} $ the Keldysh function $G_{r_{uc}}^K$ is again local in space, independent of the dimensionality, and cannot generate logarithmic terms. Focusing on the contribution of momenta $q > q_0$ it holds that
\begin{align}
 \int^\varLambda_{\gamma^{1/2}t^{1/z}} &dq q^{3-z}iG_1^K(q,t,t)=- \frac{2r_{uc}(t)}{c_K\gamma^{z/2}}\ln\left(\frac{\varLambda t^{1/z}}{\gamma^{1/2}} \right)\nonumber\\
 &=\frac{2r_{uc}(t)}{zc_K\gamma^{z/2}} \left[\ln\left(\frac{ \gamma^{z/2}}{\varLambda^zt_\gamma} \right)-\ln\left(\frac{t}{t_\gamma}\right) \right] \,.
\end{align}
Compared to the case $d < d_{uc}$ where $\epsilon > 0$, the slowly decaying correction is now logarithmically divergent both in the cut-off $\varLambda$ and at the lower boundary $q_0$. This is not surprising, because the same behavior occurs in equilibrium.
The $q$-integration of higher order terms in the expansion in $r(t)$ does not generate logarithmically divergent contributions and can therefore be neglected. Inserting the expansion $G^K_{r_{uc}} = G^K_{\text{eq}} + G^K_1$ into Eq.~\eqref{eq.1/N-for-eps=0} and keeping only the logarithmically divergent corrections, we obtain
\begin{equation}
 1= \frac{u\gamma K_d C_0}{2z\gamma^{z/2}t^{2/z}r_{uc}(t)}+\frac{uK_d}{zc_K\gamma^{z/2}}\left(\ln\left(\frac{\gamma^{z/2}}{\varLambda^z t_\gamma} \right)-\ln\left(\frac{t}{t_\gamma}\right) \right).
\end{equation}
This equation holds for all times $t>t_\gamma$ if contributions with and without $t$-dependence cancel separately. This give rise to two conditions, one for the time-dependent mass $r_{uc}(t)$ and one for the deep-quench fixed point value of $u$:
\begin{align}
 r_{uc}(t)=& \frac{\gamma c_K C_0 }{2t^{2/z}\ln(t/t_\gamma)},\label{eq:r-eps-0}\\
\label{eq:26}
 u=u^*_{uc}=&\frac{zc_K\gamma^{z/2}}{K_d\ln\bigl[ \gamma^{z/2}/(\varLambda^z t_\gamma) \bigr]}.
\end{align}
Compared to the result below $d_{uc}$ in Eqs.~\eqref{eq:fixpoint-u},~\eqref{eq:condition-for-a}, $\epsilon$ is replaced by $z/\ln(t/t_\gamma)$ in the effective mass. Since $t\gg t_\gamma$, $r_{uc}(t)$ is indeed small, justifying the earlier expansion of $G_{r_{uc}}^K$ in $r_{uc}(t)$. This additional logarithm in the self-energy correction $\delta G_{r_{uc}}^R(t,t') $ at intermediate times $t_\gamma\ll t'\ll t\ll t_q= \gamma^{z/2}q^{-z} $ implies that 
\begin{align}
 \delta G_{r_{uc}}^R(t,t')=& \int ds g^R(t,s)\frac{a \gamma}{s^{2/z}\ln(s/t_\gamma)}g^R(s,t')\nonumber\\
 \approx& \, \theta_{uc} \left\{\ln\left[\ln\left(\frac{t}{t_\gamma} \right)\right] - \ln\left[\ln\left(\frac{t'}{t_\gamma} \right)\right] \right\} \,,
\end{align}
with 
\begin{align}
  \label{eq:10}
  \theta_{uc}=-\frac{ \gamma c_K C_0 \sin(\pi/z)}{\Gamma(2/z)} \,.
\end{align}
Since the logarithmic corrections are small, we can exponentiate $\theta_{uc}$. This yields, instead of an algebraic divergence as in Eq.~\eqref{eq:scaling}, a logarithmic divergence in the scaling function
\begin{equation}
 G^R_{uc}(q,t,t')=\frac{1}{q^{2-z}\gamma^{z/2}}\left(\frac{\ln\frac{t}{t_\gamma}}{\ln\frac{t'}{t_\gamma}} \right)^{\theta_{uc}} F_{uc}^R\left(\frac{q^z t}{\gamma^{z/2}}, \frac{t}{t'} \right).
\end{equation}
The same behavior can be extended for the Keldysh-function and its scaling form at the upper critical dimension.

\section{Quench starting in the ordered phase}
\label{sec:quench-start-order}
In this section we will consider a quench with a finite initial order
parameter $\phi(t<0)=\phi_{i}$. A finite order parameter can be
achieved either by (i) a finite initial field $h_i$, which is switched off
at $t=0$ (path $B' \rightarrow C$ in Fig.~\ref{fig:2}) or by (ii) a quench in the mass-parameter with $\bar{r}_{0,i} < \bar{r}_{0,c}$ (path $B \rightarrow C$ in Fig.~\ref{fig:2}). Both quench protocols yield the same exponent $\theta$ for the retarded and the Keldysh
Green's functions. Furthermore, the exponent $\theta$ determines the
short time dynamics of $\phi(t)$ and the magnitude of the typical
cross-over time $t^*$ from prethermalized dynamics $\phi \propto t^\theta$ to relaxation $\phi \propto t^{-\beta/(\nu z)}$.\\

Starting in the symmetry-broken phase, the post-quench $1/N$-equations for a quench to the quantum critical point $r_f = 0$ read
\begin{align}
\label{eq:1}
 r(t)& = \frac{u_{f}}{2}\phi^{2}\left(t\right)+\frac{u_f}{2}\int_{q}\left(iG_{r}^{K}\left(q,t,t\right)-iG^K_{\text{eq}}(q)\right) \\
\label{eq:2}
 \phi_i\int_{-\infty}^{0}&dt'\delta\eta\left(t-t'\right) =  r(t)\phi(t)-\int_0^tds\delta\eta(t-s)\phi(s).
 \end{align}
 In the last line we dropped the $\partial_t^2\phi$ term, since we concentrate on times $t\gg t_\gamma$, where the dynamics of the system is dominated by the bath. We also assume a spatially homogeneous order parameter $\phi(t)$.

In analogy to the quench starting in the disordered phase, we expand $G^K$ in $G^K_{\text{eq}}+G^K_1$ for small $r(t)$ at long times, where $G^K_1$ is given in Eq.~\eqref{eq:GKlongtime}. This yields
\begin{align}
  r(t)=& \frac{u_{f}}{2}\phi^{2}\left(t\right)+ \frac{u_f K_d C_0}{2z\gamma^{2/z-1} t^{2/z}}\left( \frac{t}{\gamma^{z/2}}\right)^{\epsilon/z}\nonumber\\
  &+\frac{u_f}{u^*}r(t) \Bigl[ 1-\Bigl(\frac{\varLambda^zt}{\gamma^{z/2}} \Bigr)^{\epsilon/z} \Bigr] \,.
\end{align}
This equation is only fulfilled for all times $t\gg t_\gamma $ if
\begin{align}
 u_f=&u^*\\
 r(t)=&\frac{\gamma a}{ t^{2/z}}+\frac{c_K\epsilon\gamma^{z/2}\phi(t)^2}{2K_d}\left(\frac{\gamma^{z/2}}{t} \right)^{\epsilon/z} \label{eq:time-dep-mass-non-sym}
\end{align}
with  $u^*$ and $a$ given by  Eq.~\eqref{eq:fixpoint-u} and Eq.~\eqref{eq:condition-for-a}. The first equation determines the deep-quench fixed point of $u$. The value of $u^*$ is not affected by the quench direction and is idental for $A,B,B' \rightarrow C$. The second equation determines the effective, time-dependent mass. The first term in $r(t)$ is exactly the same as for the quench starting in the symmetric phase. However, the mass is now modified by an additional term proportional to $\phi^2$. This time-dependent mass can be inserted into the equation of motion~\eqref{eq:2} of $\phi$ leading to a non-linear, inhomogeneous differential equation. 

In the following two subsection we present a solution for $\phi(t)$ in the prethermalized regime and in the long time limit. The special case where the system is at the upper critical dimension $d=d_{uc}$ will be treated in section \ref{sec:order-parameter-d_uc}.
  
\subsection{Order parameter coarsening in the prethermalized regime and cross-over time scale $t^*$} 
 \label{sec:preth-regime-cross}
\begin{figure}[tb]
  \centering
  \includegraphics[width=.8\linewidth]{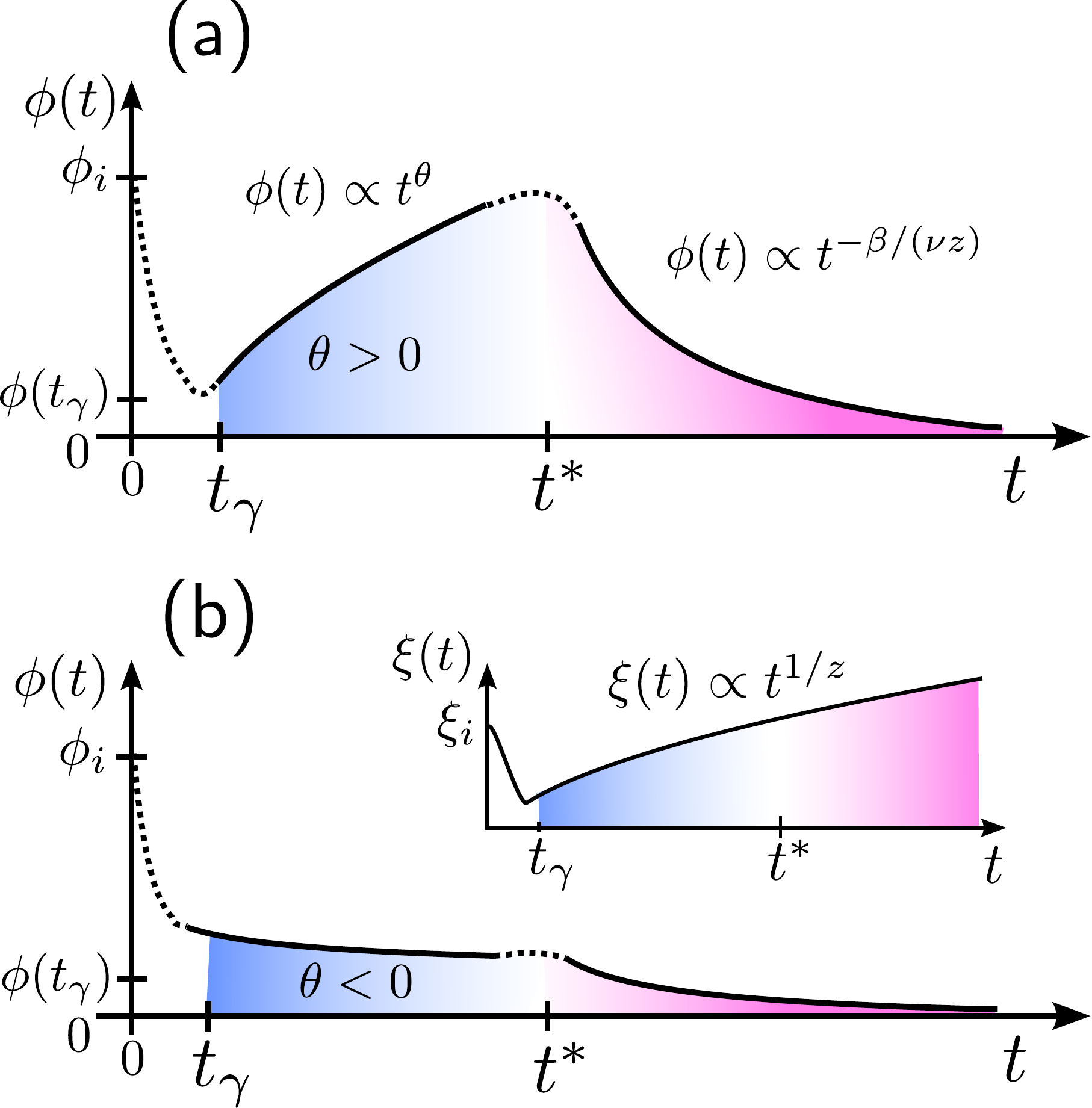}
  \caption{(Color online) Time evolution of the order parameter $\phi(t)$ following a quench from an initial state $\phi_i \neq 0$. The two panels show the cases of (a) $\theta > 0$, and (b) $\theta < 0$. First, the order parameter collapses due to the quench. After a microscopic timescale $t_\gamma$ (see also Appendix~\ref{sec:short-time-scales}), the order parameter recovers for a while $t_\gamma < t < t^*$ according to $\phi(t) \propto t^\theta$ for $\theta > 0$, before it eventually relaxes to zero for $t > t^*$ via a power law described by equilibrium exponents. For $\theta< 0$ the order parameter always decays in a power law fashion but exhibits two different exponents for $t< t^*$ and $t > t^*$. The inset in (b) shows the time dependence of the correlation length which diverges after an initial collapse in a light-cone fashion $\xi(t) \propto t^{1/z}$.  }
  \label{fig:3}
\end{figure}

 At intermediate times, one can assume that the order parameter $\phi$ is small due to the collapse of the correlations after the quench. This collapse follows \emph{a posteriori} from our solutions which also predict a collapse of the correlation length to a microscopic lengthscales right after the quench. For a fast quench, the collapse occurs on microscopic timescales. In this regime, according to Eq.~\eqref{eq:time-dep-mass-non-sym} the dominant contribution to $r(t)$ is given by
\begin{equation}
 r(t\ll t^*)\approx \frac{\gamma a}{t^{2/z}}\,,
\label{eq:preth-approx}
\end{equation}
where $t^*$ denotes the boundary of the prethermalization regime and will be defined below. 
This yields in the equation of motion~\eqref{eq:2}
\begin{align}
 \frac{\gamma a}{t^{2/z}}\phi(t)-\int_0^t ds \delta \eta(t-s)\phi(s)&=\nonumber\\
 \phi_i \int_{-\infty}^0 ds \delta \eta(t-s) .
\end{align}
One can solve this equation with the retarded Greens function $G^R$ introduced above or via Laplace transformation. Let us solve it here via Laplace-transformation. We expand $\phi(t)=\phi_0+\phi_1$ in small $a$ and obtain after a few steps
\begin{align}
 \phi_0(t)=& \phi_0(t_\gamma) \\
 \phi_1(t)=& \int_{t_\gamma}^t ds g^R(t,s)\frac{\gamma a}{s^{2/z}}\phi_0(s-t_\gamma).
\end{align}
The dominant contribution to the integral in the limit $t\gg t_\gamma$ is logarithmically divergent, yielding
\begin{equation}
 \phi(t)\approx \phi_0(t_\gamma)\left(1+\theta \log\frac{t}{t_\gamma} \right),
\end{equation}
with the same $\theta$ as in Eq.~\eqref{eq:exponent relation}. For small $\theta$ the logarithm can again be exponentiated such that 
\begin{equation}
 \phi(t)=\phi_0(t_\gamma) \left( \frac{t}{t_\gamma}\right)^\theta.
\end{equation}
For positive $\theta$ the order increases after a quench, while for negative $\theta$ it decays. Note that in both cases the order first collapses right after the quench on microscopic timescales $t_\gamma$ to a non-universal value $\phi(t_\gamma) < \phi_i$. The two cases are schematically shown in Fig.~\ref{fig:3}

In both cases the order parameter varies slowly in time compared to the algebraic decay of $\gamma a/t^{2/z}$. Therefore, independent of the sign of $\theta$ there exists a time $t^*$ where the underlying assumption of Eq.~\eqref{eq:preth-approx} that the second term in Eq.~\eqref{eq:time-dep-mass-non-sym} may be neglected compared to the first is no longer valid. This yields the crossover time $t^*$ between prethermalization and quasi-adiabatic relaxation 
\begin{equation}
 t^*= \Bigl(\frac{2 |a| K_d t_\gamma^{2\theta}}{c_K \gamma^{2/z-1+\epsilon/2}}\Bigr)^{z/(2+2z\theta-\epsilon)} \phi_0(t_\gamma)^{-\nu z/(\beta+\theta \nu z)}\label{eq:cross-over-time}\,.
\end{equation}
By performing a quench with a small amplitude such that the system was initially located close to the quantum critical point and $\phi_0(t_\gamma)$ is small, the duration of the prethermalization regime can be tuned to large timescales. Note that for $z\geq 2$ a large damping $\gamma$ increases $t^*$, while for $z<2$ a large damping shortens the prethermalized regime.

 \subsection{Quasi-adiabatic relaxation to equilibrium at long times $t\gg t^*$}
\label{sec:quasi-adiab-relax}
At long times the system thermalizes due to the external bath. For a quench right to the critical point, we expect the order parameter to eventually relax to zero. Thus, for $t \gg t^*$, where $t^*$ is the crossover time defined in Eq.~\eqref{eq:cross-over-time}, the order parameter $\phi(t)$ varies slowly in time such that one can neglect the time derivatives of $\phi(t)$ as well as the convolution term
 \begin{equation}
  \int dt' \delta \eta(t-t')\phi(t')\approx \phi(t) \int_0^\infty dt' \delta \eta(t-t')= 0 \,.
 \end{equation}
The inhomogeneous term in Eq.~\eqref{eq:eomlN-1} quickly decays to zero at long times, such that the equation of motion~\eqref{eq:2} reduces to
\begin{equation}
 r(t\gg t^*)\phi(t) = 0\,.
\label{eq:quasi-adiab-mass}
\end{equation}
A non-trivial solution of this equation requires that $r(t\gg t^*)=0$. Inserting this solution into Eq.~\eqref{eq:time-dep-mass-non-sym} yields for the dynamics of the order parameter
 \begin{align}
  \phi(t\gg t^*) = & \left( \frac{2 a K_d}{\epsilon \gamma^{2/z-1+\epsilon/2} c_K}\right)^{\frac{1}{2}} t^{-(2-\epsilon)/(2z)}\nonumber\\
  =& \left( \frac{2\gamma a \varLambda^\epsilon}{u^*\gamma^{\epsilon/2}}\right)^\beta t^{-\beta/(\nu z)}\,,
 \end{align}
where we used the large-$N$ values of the exponents $\nu$ and $\beta$. The time-dependence of $\phi$ in the long-time limit is the usual adiabatic decay and is fully determined by equilibrium exponents. This result confirms the scaling analysis in Sec.~\ref{sec:scaling-behavior}.
Like in the long time behavior of the Keldysh function, the light-cone amplitude $a$ enters as a universal prefactor.

\subsection{Dynamics of $\phi$ at $d=d_{uc}$}
\label{sec:order-parameter-d_uc}
We can extend the previous discussion to the case of a system at the upper critical dimension $d = d_{uc}$. For $\epsilon=0$ the self-consistent solution of the large-$N$ equations is given by (cf. Eqs.~\eqref{eq:r-eps-0} and~\eqref{eq:26})
\begin{align}
\label{eq:28}
u& = u^*_{uc},\\
\label{eq:27}
 r_{uc}(t)& = \frac{z }{\ln(t/t_\gamma)}\left(\frac{\gamma c_K C_0}{z t^{2/z}}+\frac{\gamma^{z/2}c_K\phi^2}{2K_d}\right).
\end{align}
In the adiabatic regime, the dynamics of $\phi$ is determined by the condition in Eq.~\eqref{eq:quasi-adiab-mass}, i.e. $r(t \gg t^*) = 0$ which yields
\begin{equation}
\label{eq:29}
 \phi(t)= \left( \frac{2K_d \gamma c_K C_0}{c_K z \gamma^{z/2}}\right)^\beta t^{-\beta/(\nu z)},
\end{equation}
where $\nu$ and $\beta$ now take their mean-field values $\beta_{\text{MF}} = \nu_{\text{MF}} = 1/2$. As expected, at $d=d_{uc}$ the long-time quasi-adiabatic dynamics is given by the mean-field values of the equilibrium exponents.

To obtain the crossover timescale $t^*$, we again compare the size of the two terms in Eq.~\eqref{eq:27}. In the prethermalized regime $t< t^*$, the  $\phi^2 $ contribution in $r_{uc}$ is much smaller than $(\gamma c_K C_0)/[t^{2/z}\ln(t/t_\gamma)] $ and can thus be neglected. In this limit, the effective mass $r_{uc}(t)$ takes the form of Eq.~\eqref{eq:r-eps-0} and it follows in analogy to the calculation in Sec.~\ref{sec:preth-regime-cross} for the dynamics of the order parameter
\begin{align}
\label{eq:30}
 \phi_{uc}(t) & \approx \phi_0\left(1+\theta_{uc}\ln\left(\ln \frac{t}{t_\gamma}\right)\right)\nonumber\\
 & \approx \phi_0\left( \ln\frac{t}{t_\gamma}\right)^{\theta_{uc}}.
\end{align}
Like in the Green's function, the additional logarithm in the effective mass leads to a even slower increase (decrease) of order parameter after the quench. 
The typical cross-over time $t^*$ is now given by
\begin{align}
\label{eq:31}
 t^*& = \left(\frac{2 |C_0| K_d}{ \gamma^{2/z-1}}\right)^{z/2}\phi_0(t_\gamma)^{-\nu z/\beta}.
\end{align}
Here, we have neglected the logarithmic contribution $\ln(t^*/t_\gamma)^{2 \theta_{uc}}$. The typical crossover time for $\epsilon=0$ is now given by the mean-field values of $\nu_{\text{MF}} = \beta_{\text{MF}} = 1/2$ and $\theta_{\text{MF}} = 0$ in the exponent of $\phi_0(t_\gamma)$. 

\section{Conclusions}
\label{sec:conclusions}
In this article we have provided a detailed analysis of the post-quench dynamics of a $\varphi^4$-model coupled to an environment, that is suddenly moved from an equilibrium state near a quantum critical point towards the quantum critical point. We have employed non-equilibrium quantum field theory in the limit of large-$N$, where $N$ is the number of field components of $\varphi$. We considered different values of the dynamic exponent $z$, which at large-$N$ is determined by the bath spectral function that we assume to be of ohmic, sub-ohmic or super-ohmic form. We investigate quenches to the quantum critical point starting from both the symmetric and the symmetry-broken phases. 

After the quench, the system is characterized by an effective, time-dependent mass $r(t)=\gamma a/t^{2/z}$ with universal amplitude $a$. The retarded and Keldysh Green's functions obey scaling $G^{R(K)} = (t/t')^\theta F^{R(K)}(q^zt/\gamma^{z/2},t/t')/(q^{2-z} \gamma^{z/2})$. Away from equilibrium, $G^{R}$ and $G^K$ depend on both time arguments $t,t'$ separately, implying interaction induced aging behavior. The singular part of the time dependence in the short time limit is characterized by a new, universal critical exponent $\theta$, which is independent of the equilibrium exponents. We have explicitly shown that one obtains the same value for $\theta$, independent of the quench direction and the initial parameters. In the long-time limit  the algebraic decay of $r(t)$ leads to a significant slowing down of thermalization. Here the value of $\theta$ enters a universal prefactor, while the aging behavior of $G^R$ and $G^K$ can be expressed by equilibrium exponents.

For a quench with a finite initial order parameter $\phi_i$, we have identified three different time regimes in the dynamics of the order parameter: (i) a non-universal regime at short times $t < t_\gamma$, (ii) a prethermalized universal regime at intermediate times $t_\gamma < t < t^*$ and (iii) a regime at long-times $t > t^*$ that is characterized by quasi-adiabatic relaxation to equilibrium. In the prethermalized regime the order parameter fulfills $\phi(t) \propto t^\theta$, where $\theta$ can be positive or negative, depending on the dynamic critical exponent $z$, which is determined by the bath spectral function. In the long-time limit the dynamics of $\phi$ is given by equilibrium exponents. The typical timescale $t^*$ separating the prethermalization from the quasi-adiabatic regime depends explicitly of the initial value of the order-parameter $\phi_i$. It diverges in the limit of a weak quench $\phi_i \rightarrow 0$. This permits tuning the prethermalization regime to extend to long time-scales, however, at the cost of a small value of the order parameter.

We also analyzed the dynamics of the order parameter $\phi(t)$ and of the retarded and correlation function $G^{R/K}$ at the upper critical dimension $d = d_{uc} = 4 - z$. Compared to the dynamics below $d_{uc}$ the power-law grow (or decay) of $\phi(t) \propto t^\theta$ in the prethermalization regime is now slowed down by the occurence of an additional logarithm $\phi_{uc}(t) \propto [\ln(t/t_\gamma)]^{\theta_{uc}} $. In the long-time limit, the dynamics of $\phi$ is given by the mean-field values of equilibrium exponents. The special $[\ln(t/t_\gamma)]^{\theta_{uc}} $ behavior arises because at the mean-field level the exponent $\theta$ vanishes. Above the upper critical dimension $d > d_{uc}$, one cannot use a scaling ansatz for $G^K$. We expect that the order parameter remains constant during the prethermalization regime, which is followed by an adiabatic decay with mean-field exponents at long times.

The unique universal dynamical behavior we report is ultimately a consquence of the collapse of the correlation function $\xi$ immediately after the quench and its recovery by critical coarsening according to $\xi(t) \propto t^{1/z}$. Due to the fast quench the system quickly falls out of equilibrium and is left in a highly exited state after the quench. Relaxation occurs due to interaction and via coupling to the external bath, which is we assume to remain at zero temperature. We observe substantial influences of quantum fluctuations and quantum coarsening on the post-quench dynamics and relaxation. Quantum fluctuations are especially important for memory effects that the system exhibits which lead to a completely different $z$ dependence of the exponent $\theta$ and of the aging prefactor of $G^{K}$ compared to the classical result. Quantum fluctuations can even result in negative exponents $\theta<0$ for super-ohmic baths. 

In the long time limit, we show that it is possible to connect the response function to the correlation function, like in the equilibrium fluctuation-dissipation theorem. This allows to introduce a non-thermal and time-dependent distribution function $n(t_a, \omega)$ which exhibits aging effects, i.\,e. it depends on the total time $t_a$ passed since the quench. At a given time, we find $n(t_a, \omega) = n_B(\omega) + \delta n(t_a, \omega)$ with $\delta n \propto \theta (|\omega| t_a)^{-2/z} [  \cos(\pi/z) + |\omega|^{-2/z} q^2 \gamma^{-1}  \sin(\pi/z) ]$. The correction $\delta n$ exhibits an algebraic frequency dependence at large frequencies which clearly distinguishes it from a thermal distribution, which shows an exponential behavior for $T> \omega$. The amplitude of the correction $\delta n$ is proportional to the short-time exponent $\theta$. The sign of $\delta n$ can either be positive or negative: while $\delta n > 0$ can be interpreted as an increased number of excitations being present compared to equilibrium, the fact that $\delta n$ can become negative implies that the density matrix of the system is not diagonal in the energy basis. The presence of quantum coherence thus defies an interpretation of $n(t_a, \omega)$ in terms of a distribution function. 

The  Young  Investigator  Group  of  P.P.O.  received  financial  support  from  the  ``Concept  for  the  Future'' of the KIT within the framework of the German Excellence Initiative.

\appendix
\section{Short-time scales}
\label{sec:short-time-scales}
Our problem is governed by several short-time scales determined by
the momentum cut of $\varLambda$ of the $\varphi^{4}-$ theory, the strength of
the damping coefficient $\gamma$ and the upper cutoff of the bath
spectrum $\omega_c$. Here we briefly discuss the hierarchy of these scales.

The large-$N$ analysis clearly shows that we will only have a solution
$r\left(t\right)\propto t^{2/z}$ if $t\gg t_{0}$ with 
\begin{equation}
t_{0}=\gamma^{z/2}/\Lambda^{z}.
\end{equation}
Otherwise the long time expansion of the upper cut off of the momentum
integration is not allowed. $t_{0}$ obeys $\Lambda^{2}=\gamma t_{0}^{-2/z}$,
i.e. it is the scale where the damping $\gamma t_0^{-2/z}$ is comparable
to the largest possible value of $k^{2}$ with momentum $k$. Clearly,
this comparison is only sensible if in fact $t_{0}>t_{c}$ with 
\begin{equation}
t_{c}=1/\omega_{c}
\end{equation}
determined by the bath cut-off. For time scales smaller than $t_{c}$,
the damping is very small and it cannot be estimated any longer via
$\gamma t^{-2/z}$ . The condition $t_{0}>t_{c}$ translates into
\begin{equation}
\gamma\omega_{c}^{2/z}>\Lambda^{2},\label{eq:c1}
\end{equation}
i.e. even the largest possible $q^{2}$ value is still smaller than
the largest possible damping term in the propagator.

Another assumption of our analysis is that we ignore the $\omega^{2}$
term relative to the damping term. This is only correct for sufficiently
low energies or, equivalently, long time scales $t\gg t_{\gamma}$
with 
\begin{equation}
t_{\gamma}=\gamma^{-z/[2(z-1)]}
\end{equation}
This result follows from $t_{\gamma}^{-2}=\gamma t_{\gamma}^{-2/z}$.
For $z>1$ discussed here, $t_{\gamma}$ can be made small if the
damping coefficient takes large values. However, comparing the ballistic
and damping term makes only sense if $t_{\gamma}>t_{c}$, which corresponds
to 
\begin{equation}
\gamma\omega_{c}^{2/z}<\omega_{c}^{2}.\label{eq:c2}
\end{equation}
The conditions Eqs.~\eqref{eq:c1} and~\eqref{eq:c2} imply that it must
hold 
\begin{equation}
\omega_{c}>\Lambda.
\end{equation}
One can also introduce the time scale 
\begin{equation}
t_{\Lambda}=1/\Lambda
\end{equation}
which is always larger than $t_{c}.$ Thus, the shortest scale is
always $t_{c}$.

Our prethermalized regime can only start after ${\rm max}\left(t_{0},t_{\gamma},t_{\Lambda}\right)$.
One finds that no matter what the ratio of $t_{0}$ and $t_{\gamma}$,
$t_{\Lambda}$ is always between these two time scales and can therefore
never be the largest. As for the ratio between $t_{0}$ and $t_{\gamma}$,
it seems plausible to request $t_{\gamma}\gg t_{\Lambda}$ since $t_{\gamma}$
is the only scale that doesn't depend on a cut-off. In this case we
have the following hierarchy of scales 
\begin{equation}
\omega_{c}\gg\Lambda^{z}/\gamma^{z/2}\gg\Lambda\gg t_{\gamma}^{-1}
\end{equation}
and universal prethermalized behavior will set in after $t_{\gamma}$.
Thus, an important condition for our analysis is that the damping
due to the bath is sufficiently large such that there is a wide regime $t\gg t_\gamma$, where damping dominates  the dynamics of the system.

\section{Large-$N$ expansion after a quantum quench}
\label{sec:large-n-expansion}
In this appendix we summarize the derivation of the large-$N$ equations
for the out-of-equilibrium dynamics after a sudden quench. After having
integrated out the bath degrees of freedom the action of the collective
order parameter field in Eq.~\eqref{eq:23} is given in terms of the Matsubara, quantum
and classical field variables as 
\begin{equation}
S\left[\boldsymbol{\varphi}\right]=S_{M}\left[\boldsymbol{\varphi}_{M}\right]+S_{K}\left[\boldsymbol{\varphi}_{c},\boldsymbol{\varphi}_{q}\right]+S_{C}\left[\boldsymbol{\varphi}_{M},\boldsymbol{\varphi}_{q}\right].
\end{equation}
The first term is up to a trivial factor $i$ the usual Matsubara
action in equilibrium
\begin{eqnarray}
S_{M}\left[\boldsymbol{\varphi}_{M}\right] & = & \frac{i}{2}\int_{x,\tau\tau'}\boldsymbol{\varphi}_{M}\left(\tau\right) (g_i^{M})^{-1}\left(\tau-\tau'\right)\boldsymbol{\varphi}\left(_{M}\tau'\right)\nonumber \\
 & + & i\frac{u_{i}}{4N}\int_{x,\tau}\left(\boldsymbol{\varphi}_{M}\left(\tau\right)\cdot\boldsymbol{\varphi}_{M}\left(\tau\right)\right)^{2},
\end{eqnarray}
where $g_i^{M}\left(\tau-\tau'\right)$ follows from Eq.~\eqref{eq:G0Matsubara}
after Fourier transformation. We use the notation $\int_{\tau}\equiv\int_{0}^{\beta}d\tau$
with usual imaginary time $\tau$ defined via $t=i\tau$. For the
time integrals along the real part we use $\int_{t}\equiv\int_{0}^{\infty}dt$.
The two segments on the real axis are given as:
\begin{align}
S_{K}& \left[\boldsymbol{\varphi}_{c}, \boldsymbol{\varphi}_{q}\right] =  \frac{1}{2}\int_{x,tt'}\left(\begin{array}{c}
\boldsymbol{\varphi}_{c}\left(t\right)\\
\boldsymbol{\varphi}_{q}\left(t\right)
\end{array}\right)^{T}{\cal G}_{0}^{-1}\left(t,t'\right)\left(\begin{array}{c}
\boldsymbol{\varphi}_{c}\left(t'\right)\\
\boldsymbol{\varphi}_{q}\left(t'\right)
\end{array}\right)\nonumber \\
 & - \frac{u_{f}}{2N}\int_{x,t}\left(\boldsymbol{\varphi}_{c}^{2}\left(t\right)+\boldsymbol{\varphi}_{q}^{2}\left(t\right)\right)\boldsymbol{\varphi}_{c}\left(t\right)\cdot\boldsymbol{\varphi}_{q}\left(t\right)
\end{align}
with matrix propagator:
\begin{equation}
{\cal G}_{0}=\left(\begin{array}{cc}
g_{f}^{K} & g_{f}^{R}\\
g_{f}^{A} & 0
\end{array}\right).
\end{equation}
The inverse is given as 
\begin{equation}
{\cal G}_{0}^{-1}=\left(\begin{array}{cc}
0 & (g_{f}^{A})^{-1}\\
(g_{f}^{R})^{-1} & M_{0}
\end{array}\right),
\end{equation}
where $M_{0}\equiv \left(\mathcal{G}_{0}^{-1}\right)^{K}=-(g_{f}^{R})^{-1} g_{f}^{K} (g_{f}^{A})^{-1}$.
Since we integrated out the bath degrees of freedom the retarded Green's
function is determined by Eq.~\eqref{eq:Goretarded} while
\begin{equation}
M_{0}\left(t,t'\right)=-\nu\left(t-t'\right).
\end{equation}
The fluctuation-dissipation theorem of the bath can be expressed by
the Fourier-transforms 
\begin{equation}
\nu(\omega)=\coth\left(\frac{\omega}{2T}\right)\left(\eta(\omega)-\eta^{*}(\omega)\right).
\end{equation}
The coupling between the two parts of the contour is
\begin{equation}
S_{C}\left[\boldsymbol{\varphi}_{M},\boldsymbol{\varphi}_{q}\right]=i\int_{t,\tau}\boldsymbol{\varphi}_{M}\left(\tau\right)\tilde{\eta}\left(i\tau,t\right)\boldsymbol{\varphi}_{q}\left(t\right),
\end{equation}
where $\tilde{\eta}\left(t\right)=\sqrt{2}\left(\eta^{<}\left(t\right)-\eta^{<}\left(-t\right)\right)$
and $\eta^{\gtrless}=\frac{1}{2}(\nu\pm\eta\mp\eta^{*})$.

Since $\eta_{0}\equiv\eta\left(\omega=0\right)$ is finite, we have
$\eta\left(t-t'\right)=\eta_{0}\delta\left(t-t'\right)+\delta\eta\left(t-t'\right)$,
where $\delta\eta\left(t-t'\right)$ is purely non-local. This yields with $\bar{r}_{0,f}=r_{0,f} - \eta_{0}$ that 
\begin{eqnarray}
(g_{f}^{R})^{-1}\left(t,t'\right) & = & -\left(\partial_{t}^{2}+\bar{r}_{0,f}-\nabla^{2}\right)\delta\left(t-t'\right)\nonumber \\
 & + & \delta\eta\left(t-t'\right).
\end{eqnarray}
Since $\nu\left(0\right)=0$, $\nu\left(t-t'\right)$ is purely non-local
in time. Using $\eta\left(\omega=0\right)=\eta_{M}\left(i\omega_{n}=0\right)$
we can absorb the same zero frequency part on the imaginary axis:
$\bar{r}_{0,i}=r_{0,i}-\eta_{0}$, such that 
\begin{eqnarray}
(g_{f}^{M})^{-1}\left(\tau,\tau'\right) & = & \left(\partial_{\tau}^{2}-\bar{r}_{0,i}+\nabla^{2}\right)\delta\left(\tau-\tau'\right)\nonumber \\
 & + & \delta\eta_{M}\left(\tau-\tau'\right).
\end{eqnarray}

Next we perform the large-$N$ expansion and follow closely the procedure
in chapter 30 of Ref.~\onlinecite{zinn-justin_QFT}. Important degrees of freedom are obviously $\rho_{\pm}=\frac{1}{N}\boldsymbol{\varphi}_{\pm}^{2}$
and $\rho_{M}=\frac{1}{N}\boldsymbol{\varphi}_{M}^{2}$ which yields
in the case of the round trip segment 
\begin{eqnarray}
\rho_{c} & = & \frac{1}{\sqrt{2}}\left(\rho_{+}+\rho_{-}\right)=\frac{1}{\sqrt{2}N}\left(\boldsymbol{\varphi}_{c}^{2}+\boldsymbol{\varphi}_{q}^{2}\right)\nonumber \\
\rho_{q} & = & \frac{1}{\sqrt{2}}\left(\rho_{+}-\rho_{-}\right)=\frac{\sqrt{2}}{N}\boldsymbol{\varphi}_{c}\cdot\boldsymbol{\varphi}_{q}.
\end{eqnarray}
The interaction term of the action along the real axis part of the
contour is then given by
\begin{eqnarray}
S_{K,{\rm int}} & = & -\frac{u_{f}}{2N}\int_{x,t}\left(\boldsymbol{\varphi}_{c}^{2}\left(t\right)+\boldsymbol{\varphi}_{q}^{2}\left(t\right)\right)\boldsymbol{\varphi}_{c}\left(t\right)\cdot\boldsymbol{\varphi}_{q}\left(t\right)\nonumber \\
 & = & -\frac{u_{f}N}{2}\int_{x,t}\rho_{c}\left(t\right)\rho_{q}\left(t\right).
\end{eqnarray}
In order to enforce the above constraints, we introduce 
\begin{eqnarray}
1 & = & \int D\rho\prod_{x,t}\delta\left(\frac{1}{\sqrt{2}N}\left(\boldsymbol{\varphi}_{c}^{2}+\boldsymbol{\varphi}_{q}^{2}\right)-\rho_{c}\right)\nonumber \\
 & \times & \prod_{x,t}\delta\left(\frac{\sqrt{2}}{N}\boldsymbol{\varphi}_{c}\cdot\boldsymbol{\varphi}_{q}-\rho_{q}\right)\prod_{x,\tau}\delta\left(\frac{1}{N}\boldsymbol{\varphi}_{M}^{2}-\rho_{M}\right)\nonumber \\
 & \propto & \int D\rho Dre^{-\frac{i}{2}\int_{x,t}r_{c}\left(t\right)\left(2\boldsymbol{\varphi}_{c}\cdot\boldsymbol{\varphi}_{q}-\sqrt{2}N\rho_{q}\right)}\nonumber \\
 & \times & e^{-\frac{i}{2}\int_{x,t}r_{q}\left(t\right)\left(\boldsymbol{\varphi}_{c}^{2}+\boldsymbol{\varphi}_{q}^{2}-\sqrt{2}N\rho_{c}\right)}\nonumber \\
 & \times & e^{-\frac{1}{2}\int_{x,\tau}r_{M}\left(t\right)\left(\boldsymbol{\varphi}_{M}^{2}-N\rho_{M}\right)}
\end{eqnarray}
where $\int D\rho$ $ $is the short hand for $\int D\rho_{c}\int D\rho_{q}\int D\rho_{M}$
and similar for other degrees of freedom. The integration contour
for the $r_{c,q}$-integrations is along the real axis, while the
$r_{M}$-integral is performed along the imaginary axis. It follows
for the generating functional 
\[
Z=\int D\boldsymbol{\varphi}D\rho Dre^{i\left(S_{M}+S_{K}+S_{C}\right)}
\]
with unchanged coupling term $S_{C}$ and
\begin{eqnarray*}
S_{M} & = & \frac{i}{2}\int_{x,\tau\tau'}\boldsymbol{\varphi}_{M}\left(\tau\right) (G_{r}^{M})^{-1}\left(\tau,\tau'\right)\boldsymbol{\varphi}_{M}\left(\tau'\right)\\
 & + & i\frac{N}{2}\int_{x,\tau}\left(\frac{u_{i}}{2}\rho_{M}^{2}\left(\tau\right)+\left(\bar{r}_{0,i}-r_{M}\left(\tau\right)\right)\rho_{M}\left(\tau\right)\right)
\end{eqnarray*}
as well as 
\begin{eqnarray}
S_{K} & = & \frac{1}{2}\int_{x,t,t'}\left(\begin{array}{c}
\boldsymbol{\varphi}_{c}\left(t\right)\\
\boldsymbol{\varphi}_{q}\left(t\right)
\end{array}\right)^{T}{\cal G}_{r}^{-1}\left(t,t'\right)\left(\begin{array}{c}
\boldsymbol{\varphi}_{c}\left(t'\right)\\
\boldsymbol{\varphi}_{q}\left(t'\right)
\end{array}\right)\nonumber \\
 & - & \frac{N}{2}\int_{x,t}\left(\sqrt{2}\left(\bar{r}_{0,f}-r_{c}\left(t\right)\right)\rho_{q}\left(t\right)+u_{f}\rho_{q}\left(t\right)\rho_{c}\left(t\right)\right)\nonumber \\
 & - & \frac{N}{2}\int_{x,t}\sqrt{2}r_{q}\left(t\right)\rho_{c}\left(t\right).
\end{eqnarray}
The modified propagators are given as 
\begin{eqnarray}
(G_{r}^{M})^{-1}\left(\tau,\tau'\right) & = & \left(-\partial_{\tau}^{2}+r_{M}\left(\tau\right)-\nabla^{2}\right)\delta\left(\tau-\tau'\right)\nonumber \\
 & + & \delta\eta_{M}\left(\tau-\tau'\right)
\end{eqnarray}
on the imaginary axis and 
\begin{equation}
{\cal G}_{r}^{-1}\left(t,t'\right)=\left(\begin{array}{cc}
0 & (G_{r}^{A})^{-1}\\
(G_{r}^{R})^{-1} & M_{0}
\end{array}\right)_{t,t'}-r_{q}\left(t\right)\delta\left(t-t'\right) \openone
\end{equation}
with modified retarded function 
\begin{equation}
(G_r^{R})^{-1}\left(t,t'\right)=-\left(\partial_{t}^{2}+r_{c}\left(t\right)-\nabla^{2}\right)\delta\left(t-t'\right)+\delta\eta\left(t-t'\right)
\end{equation}
on the real axis and $2\times2$ unit matrix $\openone$. Thus,
the bare masses have been replaced by in general time dependent terms
that corresponds to become the usual self energy corrections of the
large-$N$ theory.

We express the vectors $\boldsymbol{\varphi}_{c,q,M}$ in terms of
the component $\phi_{c,q,M}$ along the direction of the field $\mathbf{h}_{i,m}=\left(h_{i,m},0,\dots,0\right)$
and the $N-1$ components $\boldsymbol{\pi}_{c,q,M}$ orthogonal to
it. 
\begin{equation}
\boldsymbol{\varphi}_{c,q,M}=\left(\sqrt{N}\phi_{c,q,M},\boldsymbol{\pi}_{c,q,M}\right).
\end{equation}
We first integrate over the transverse fields $\boldsymbol{\pi}_{M}$
along the imaginary contour and obtain 
\begin{eqnarray*}
S & = & S_{K}+\frac{iN}{2}\int_{x,\tau\tau'}\phi_{M}\left(\tau\right) (G_{r}^{M})^{-1}\left(\tau,\tau'\right)\phi_{M}\left(\tau'\right)\\
 & + & iN\int_{x,\tau t}\phi_{M}\left(\tau\right)\tilde{\eta}\left(i\tau,t\right)\phi_{q}\left(t\right)\\
 & + & \frac{i}{2}\left(N-1\right){\rm tr}\log (G_{r}^{M})^{-1}-\frac{i}{2}\int_{x,tt'}\boldsymbol{\pi}_{q}\left(t\right)\mu\left(t,t'\right)\boldsymbol{\pi}_{q}\left(t'\right)\\
 & + & i\frac{N}{2}\int_{x,\tau}\left(\frac{u_{i}}{2}\rho_{M}^{2}\left(\tau\right)+\bar{r}_{0,i}\rho_{M}\left(\tau\right)-r_{M}\left(\tau\right)\rho_{M}\left(\tau\right)\right).
\end{eqnarray*}
where 
\begin{equation}
\mu\left(t,t'\right)=\int_{\tau\tau'}\tilde{\eta}\left(t,i\tau\right)G_{r}^{M}\left(\tau,\tau'\right)\tilde{\eta}\left(i\tau',t'\right).
\end{equation}
The term $\int\boldsymbol{\pi}_{q}\mu\boldsymbol{\pi}_{q}$ can be
absorbed into a redefinition of the bare inverse Keldysh function
\begin{equation}
M_{0}\left(t,t'\right)\rightarrow M\left(t,t'\right)=-\nu\left(t-t'\right)-\mu\left(t,t'\right).
\end{equation}
This is a crucial aspect of the theory demonstrating that the bare post-quench correlation function is affected by the pre-quench behavior due to memory effects of the bath.

Next, we integrate over the transverse fields $\boldsymbol{\pi}_{c,q}$ and obtain 
\begin{eqnarray}
S & = & \frac{N}{2}\int_{x,t}\left(\begin{array}{c}
\phi_{c}\left(t\right)\\
\phi_{q}\left(t\right)
\end{array}\right)^{T}{\cal G}_r^{-1}\left(t,t'\right)\left(\begin{array}{c}
\phi_{c}\left(t'\right)\\
\phi_{q}\left(t'\right)
\end{array}\right)  \\
 & + & \frac{iN}{2}\int_{x,\tau\tau'}\phi_{M}\left(\tau\right) (G_{r}^{M})^{-1}\left(\tau,\tau'\right)\phi_{M}\left(\tau'\right)\nonumber \\
 & + & iN\int_{x,\tau t}\phi_{M}\left(\tau\right)\tilde{\eta}\left(i\tau,t\right)\phi_{q}\left(t\right)\nonumber \\
 & - & \frac{N}{2}\int_{x,t}\left(\sqrt{2}\left(\bar{r}_{0,f}-r_{c}\left(t\right)\right)\rho_{q}\left(t\right)+u_{f}\rho_{q}\left(t\right)\rho_{c}\left(t\right)\right)\nonumber \\
 & + & \frac{N}{2}\int_{x,t}\sqrt{2}r_{q}\left(t\right)\rho_{c}\left(t\right)\nonumber \\
 & + & i\frac{N}{2}\int_{x,\tau}\left(\bar{r}_{0,i}\rho_{M}\left(\tau\right)+\frac{u_{i}}{2}\rho_{M}^{2}\left(\tau\right)-r_{M}\left(\tau\right)\rho_{M}\left(\tau\right)\right).\nonumber \\
 & + & \frac{i}{2}\left(N-1\right){\rm tr}\log{\cal G}_{r}^{-1}+\frac{i}{2}\left(N-1\right){\rm tr}\log (G_{r}^{M})^{-1} \nonumber \,.
\end{eqnarray}
The action is of order $N$, allowing us to perform a saddle point
analysis. Minimizing the action with respect to $\phi$, $\rho$,
and $r$ yields the nine saddle point equations. Minimizing with respect
to the fields on the Matsubara segment of the contour yields
\begin{eqnarray}
h_{M}\left(\tau\right) & = & \int_{\tau'} (G_{r}^{M})^{-1}\left(\tau,\tau'\right)\phi_{M}\left(\tau'\right)\nonumber \\
 & + & \int_{t}\tilde{\eta}\left(i\tau,t\right)\phi_{q}\left(t\right)\nonumber \\
\rho_{M}\left(\tau\right) & = & \phi_{M}^{2}\left(\tau\right)+\frac{i}{2}\frac{\delta{\rm tr}\log{\cal G}_{r}^{-1}}{\delta r_{M}\left(\tau\right)}+\frac{\delta{\rm tr}\log (G_{r}^{M})^{-1}}{\delta r_{M}\left(\tau\right)}\nonumber \\
r_{M}\left(\tau\right) & = & \bar{r}_{0,i}+u_{i}\rho_{M}\left(t\right)
\end{eqnarray}
The minimization with respect to the quantum field on the Keldysh
contour yields
\begin{eqnarray}
h_{c}\left(t\right) & = & -r_{q}\left(t\right)\phi_{q}\left(t\right)+\int dt' (G_{r}^{R})^{-1}\left(t,t'\right)\phi_{c}\left(t'\right)\nonumber \\
 & + & \int dt'\left(G_{r}^{-1}\right)^{K}\left(t,t'\right)\phi_{q}\left(t'\right)+\int_{\tau}\phi_{M}\left(\tau\right)\tilde{\eta}\left(i\tau,t\right)\nonumber \\
\rho_{c}\left(t\right) & = & \frac{1}{\sqrt{2}}\left(\phi_{c}^{2}\left(t\right)+\phi_{q}^{2}\left(t\right)\right)-\frac{i}{\sqrt{2}}\frac{\delta{\rm tr}\log{\cal G}_{r}^{-1}}{\delta r_{q}\left(t\right)}\nonumber \\
r_{c}\left(t\right) & = & \bar{r}_{0,f}+\frac{u_{f}}{\sqrt{2}}\rho_{c}\left(t\right).
\end{eqnarray}
Finally, from the minimization with respect to the classical fields
on the Keldysh contour follows that 
\begin{eqnarray}
h_{q}\left(t\right) & = & -r_{q}\left(t\right)\phi_{c}\left(t\right)+\int_{t'} (G_r^{A})^{-1}\left(t,t'\right)\phi_{q}\left(t'\right)\nonumber \\
\rho_{q}\left(t\right) & = & \frac{1}{\sqrt{2}}\phi_{c}\left(t\right)\phi_{q}\left(t\right)-\frac{i}{\sqrt{2}}\frac{\delta{\rm tr}\log{\cal G}_{r}^{-1}}{\delta r_{c}\left(t\right)}\nonumber \\
r_{q}\left(t\right) & = & \frac{u}{\sqrt{2}}\rho_{q}\left(t\right).
\end{eqnarray}
The natural saddle point solution for the quantum components respecting
causality is $r_{q}=\rho_{q}=\phi_{q}=0$ for $h_{q}=0$. We checked that this trivial
solution is the only solution by an explicit analyses of the Heisenberg equation of motion of the order parameter and the propagators. Performing the
functional derivatives and evaluating them at vanishing quantum fields yields
\begin{eqnarray}
\frac{\delta{\rm tr}\log{\cal G}_r^{-1}}{\delta r_{c}\left(t\right)} & = & {\rm tr}\left({\cal G}_r\frac{\delta{\cal G}_r^{-1}}{\delta r_{c}\left(t\right)}\right)\nonumber \\
 & = & -G_r^{R}\left(t,t\right)-G_r^{A}\left(t,t\right)=0\nonumber \\
\frac{\delta{\rm tr}\log{\cal G}_r^{-1}}{\delta r_{q}\left(t\right)} & = & {\rm tr}\left({\cal G}_r\frac{\delta{\cal G}_r^{-1}}{\delta r_{q}\left(t\right)}\right)=-G_r^{K}\left(t,t\right).\nonumber \\
\frac{\delta{\rm tr}\log (G_{r}^{M})^{-1}}{\delta r_{M}\left(\tau\right)} & = & (G_{r}^{M})^{-1}\left(\tau,\tau\right)\nonumber \\
\frac{\delta{\rm tr}\log{\cal G}_r^{-1}}{\delta r_{M}\left(\tau\right)} & = & 0.
\end{eqnarray}
We consider time-independent fields $h_{M}\left(\tau\right)=h_{i}$
and $h_{c}\left(t\right)=h_{f}$ before and after the quench, respectively.
Since the imaginary time evolution prior to the quench is in equilibrium
we obtain time independent solutions for $r_{M}=r_{i}$ and $\phi_{M}=\phi_{i}$,
yielding the usual equilibrium version of the large-$N$ equations.
Eliminating $\rho_{M}$ yields
\begin{eqnarray}
h_{M} & = & r_{M}\phi_{M}\nonumber \\
r_{M} & = & \bar{r}_{0,i}+\frac{u_{i}}{2}\phi_{M}^{2}+u_{i}G_{r}^{M}\left(x,\tau;x,\tau\right).
\end{eqnarray}
In the last step we reintroduced spatial coordinates. The non-equilibrium
large-$N$ equations of the classical components $r_{c}$, $\rho_{c}$,
and $\phi_{c}$ are nontrivial and time dependent. We obtain after the elimination
of $\rho$:
\begin{eqnarray}
h_{c} & = & \int dt' (G_{r}^{R})^{-1}\left(t,t'\right)\phi_{c}\left(t'\right)+\phi_{M}\int_{\tau}\tilde{\eta}\left(i\tau,t\right)\nonumber \\
r\left(t\right) & = & \bar{r}_{0,f}+\frac{u_{f}}{2}\phi_{c}^{2}\left(t\right)+\frac{u}{2}iG_{r}^{K}\left(x,t;x,t\right).
\end{eqnarray}
These are the self-consistent large-$N$ equations given in the main
part of the paper. There we only used the more physically motivated
notation $h_{i}=h_{M}$, $\phi_{i}=\phi_{M},$ and $r_{i}=r_{M}$
to indicate that the variables on the Matsubara branch refer to the
initial field, order-parameter, and renormalized mass, respectively.
Similarly we have $h_{f}=h_{c}, \phi\left(t\right)=\phi_{c}\left(t\right),$
and $r\left(t\right)=r_{c}\left(t\right)$ for the corresponding variables
after the quench.

\section{The fixed mass post-quench Green's functions}\label{app:post-quench G}

In this subsection we summarize the main steps in the derivation of
the post-quench retarded and Keldysh function of the system with constant
masses $r$. The knowledge of this propagator is essential for any
development of a perturbative approach to includes interactions.

In order to determine the bare propagators, we start from the Heisenberg
equation of motion of the field operator after the quench: 
\begin{eqnarray}
\left(\partial_{t}^{2}+r_{0,f}+q^{2}\right)\boldsymbol{\varphi}\left(q,t\right) & = & \int_{0}^{\infty}ds\eta\left(t-s\right)\boldsymbol{\varphi}\left(q,s\right)\nonumber \\
 & + & \boldsymbol{\Xi}\left(q,t\right)+\mathbf{h}\left(q,t\right)\label{eq:eom}
\end{eqnarray}
with external field $\mathbf{h}\left(q,t\right)$ that is coupled
to the order parameter. The source operator $\boldsymbol{\Xi}(q,t)$
is given by 
\begin{equation}
\boldsymbol{\Xi}(q,t)=-\int_{-\infty}^{0}ds\eta(t-s)\boldsymbol{\varphi}(q,s).
\end{equation}
It is useful to express $\boldsymbol{\Xi}(t)$ in therms of the initial
bath-operators $\mathbf{X}_{j}^{0}=\mathbf{X}_{j}\left(q,t=0\right)$
and $\mathbf{P}_{j}^{0}=\mathbf{P}_{j}\left(q,t=0\right)$: 
\[
\boldsymbol{\Xi}\left(q,t\right)=-\sum_{j}c_{j}\left(\mathbf{X}_{j}^{0}\left(q\right)\cos\left(\Omega_{j}t\right)+\frac{1}{\Omega_{j}}\mathbf{P}_{j}^{0}\left(q\right)\sin\left(\Omega_{j}t\right)\right)
\]
We solve the Heisenberg equation of motion via Laplace transform 
\begin{equation}
\boldsymbol{\varphi}\left(q,\omega\right)=\int_{0}^{\infty}dte^{i\left(\omega+i0^{+}\right)t}\boldsymbol{\varphi}\left(q,t\right)
\end{equation}
with boundary condition $\boldsymbol{\varphi}\left(q,t=0\right)=\boldsymbol{\varphi}_{0}\left(q\right)$
and $\partial_{t}\boldsymbol{\varphi}\left(q,t=0\right)=\boldsymbol{\pi}_{0}\left(q\right)$.
It follows 
\begin{equation}
\boldsymbol{\varphi}\left(q,\omega\right)=\boldsymbol{F}\left(q,\omega\right)g_{f}^{R}\left(q,\omega\right)\label{eq:Laplace OP}
\end{equation}
with force-operator 
\begin{equation}
\boldsymbol{F}\left(q,\omega\right)=\boldsymbol{\pi}_{0}\left(q\right)-i\omega\boldsymbol{\varphi}_{0}\left(q\right)+\mathbf{\boldsymbol{\Xi}}\left(q,\omega\right)+\mathbf{h}\left(q,\omega\right)\label{eq:force}
\end{equation}
and bare retarded post-quench Green's function 
\begin{equation}
g_{f}^{R}\left(q,\omega\right)=\frac{1}{\omega^{2}-r_{0,f}-q^{2}+\eta\left(\omega\right)}.\label{eq:GR0}
\end{equation}
To check that this is indeed the correct result for the bare retarded
Green's function one expresses the back transform $\boldsymbol{\varphi}\left(q,t\right)$
of $\boldsymbol{\varphi}\left(q,\omega\right)$ in terms of $g_{f}^{R}\left(k,\omega\right)$:
\begin{eqnarray}
\boldsymbol{\varphi}\left(q,t\right) & = & \left(\boldsymbol{\pi}_{0}\left(q\right)+\boldsymbol{\varphi}_{0}\left(q\right)\partial_{t}\right)g_{f}^{R}\left(q,t\right).\label{eq:Laplace sol}\\
 & + & \int_{0}^{\infty}\left(\mathbf{\boldsymbol{\Xi}}\left(q,s\right)+\mathbf{h}\left(q,s\right)\right)g_{f}^{R}\left(k,t-s\right)ds\nonumber 
\end{eqnarray}
Inserting this result into the definition Eq.~\eqref{eq:defGR} of the
retarded function does indeed yield Eq.~\eqref{eq:GR0} for zero external
field $\mathbf{h}\left(q,t\right)$. Thus, for the bare retarded post-quench
Green's function follows the same result as in equilibrium, which
only depends on the difference between the two time arguments. As
we will see below, this behavior of the retarded function will not
carry over to case where we include interactions.

Next we consider the bare post-quench Keldysh function $g_{f}^{K}\left(q,t,t'\right)$.
This function will depend on both time scales already for a non-interacting
system. We determine the $G_{f}^{K}\left(q,t,t'\right)$ using the
same approach as for the retarded function, i.e. insert the solution
Eq.~\eqref{eq:Laplace sol} into the definition Eq.~\eqref{eq:defGK} of
the Keldysh function. Here, it is convenient to consider the double
Laplace transform: 
\begin{equation}
g_{f}^{K}\left(q,\omega,\omega'\right)=\int_{0}^{\infty}dtdt'g_{f}^{K}\left(q,t,t'\right)e^{i\left(\omega_{+}t+\omega'_{+}t'\right)},
\end{equation}
with $\omega_{+}=\omega+i0^{+}$. From our solution Eq.~\eqref{eq:Laplace OP}
follows 
\begin{equation}
g_{f}^{K}\left(q,\omega,\omega'\right)=M\left(q,\omega,\omega'\right)g_{f}^{R}\left(q,\omega\right)g_{f}^{R}\left(q,\omega'\right),\label{eq:GK memory}
\end{equation}
with memory function 
\begin{equation}
M\left(q,\omega,\omega'\right)=-i\left\langle \left[F_{a}\left(q,\omega\right),F_{a}\left(-q,\omega'\right)\right]_{+}\right\rangle ,
\end{equation}
given by the force-force correlation function. $M\left(q,\omega,\omega'\right)$
can be obtained from the expectation values of $\boldsymbol{F}\left(q,\omega\right)$
as given in Eq.~\eqref{eq:force}. To proceed we note that the operators
that enter the memory function are $\boldsymbol{\varphi}_{0}\left(q\right)$,
$\mathbf{X}_{j}^{0}\left(q\right)$ etc., i.e. they are for system
and bath variables at $t=0$. Thus, all relevant expectation values
can be evaluated in equilibrium prior to the quench. The direct evaluation
of these expectation values is straightforward but somewhat tedious.
In particular one finds that numerous expectation values diverge in
the limit of an infinite cut-off $\omega_{c}$, divergences that
 cancel if one combines all terms that contribute to the
memory function. To avoid these complications we present a significantly
easier approach to this problem. We explicitly checked that both approaches
lead to the same result. We stress again that all expectation values
that enter the memory function can equally be determined prior to
the quench when the system is still in equilibrium. We therefore consider
a system without quench and with initial Hamiltonian $H_{i}$ for
all times. The Keldysh function of this equilibrium system should
of course have the same formal structure as Eq.~\eqref{eq:GK memory},
i.e.: 
\begin{equation}
g_{{i}}^{K}\left(q,\omega,\omega'\right)=M\left(q,\omega,\omega'\right)g_{i}^{R}\left(q,\omega\right)g_{i}^{R}\left(q,\omega'\right)
\end{equation}
The subscript $i$ indicates that we are considering a system in equilibrium
that is governed by the Hamiltonian $H_{i}$ with bare mass $r_{0,i}$. The key insight is that
the memory function $M\left(q,\omega,\omega'\right)$ must be the
same function as in Eq.~\eqref{eq:GK memory} as we have to determine
the expectation value of the same operators $\mathbf{F}\left(q,\omega\right)$
with respect to the same state. 

In equilibrium, the Keldysh function only depends on the difference of the time arguments. For the double Laplace transform this implies 
\begin{equation}
g_{i}^{K}\left(q,\omega,\omega'\right)=i\frac{g_{i}^{K}\left(q,\omega\right)+g_{i}^{K}\left(q,\omega'\right)}{\omega+\omega'+i0^{+}},
\end{equation}
where $g_{i}^{K}\left(q,\omega\right)$ is the Laplace transform
of the bare Keldysh function of the initial state in equilibrium:
\begin{equation}
g_{i}^{K}\left(q,\omega\right)=\int_{0}^{\infty}dte^{i(\omega+i0^{+})t}g_{i}^{K}\left(q,t\right).
\end{equation}
In equilibrium we can then use the fluctuation-dissipation relation
\begin{equation}
g_{i}^{K}\left(q,t\right)=i\int_{-\infty}^{\infty}\frac{d\epsilon}{\pi}e^{-i\epsilon t}\coth\left(\frac{\epsilon}{2T}\right){\rm Im}g_{i}^{R}\left(q,\epsilon\right)
\end{equation}
to determine this function. Thus, we obtain for the memory function
\begin{eqnarray}
M\left(q,\omega,\omega'\right) & = & i\frac{g_{i}^{K}\left(q,\omega\right)+g_{i}^{K}\left(q,\omega'\right)}{\omega+\omega'+i0^{+}}\nonumber \\
 & \times & g_{i}^{R}\left(q,\omega\right)^{-1} g_{i}^{R}\left(q,\omega'\right)^{-1}.\label{eq:memory}
\end{eqnarray}
As mentioned, we obtained the same result using a straightforward
evaluation of the definition of the Keldysh function, inserting the
solution Eq.~\eqref{eq:Laplace sol}, and evaluating all expectation
values explicitly. As limiting cases, Eq.~\eqref{eq:memory} includes
the solution of Ref.~\onlinecite{Janssen-ZPhysB-1989} for a classical system
with ohmic bath and Ref.~\onlinecite{1742-5468-2012-01-P01014} for a classical system with
colored noise. In addition it reproduces the findings of Ref.~\onlinecite{PhysRevB.81.134305} for a non-interacting quantum system without bath.

As discussed in the context of the scaling behavior, the limit where
we consider large distances $\delta r_{0i}$ from the critical point
prior to the quench is of particular interest. In this deep quench
limit follows: 
\begin{equation}
M(q,\omega,\omega')=i\frac{\coth(\frac{\omega}{2T})\delta\eta(\omega)+\coth(\frac{\omega}{2T})\delta\eta(\omega')}{\omega+\omega'+i0^{+}}.\label{eq:memodq}
\end{equation}
For times $t,t'>t_{i}=(\gamma/\delta r_i)^{z/2}$ the memory function is always governed by
this deep-quench limit, even if $\delta r_{i}$ is nominally not large. This enables us to introduce a scaling form for $G^K$ and to expect universal behavior after the quench.
The justification to use the deep-quench limit in this paper is however not, that we concentrate one time-scales larger than $t_i$, but rather that the even for a small quench the spatial correlation length collapses and the bar Green's function become completely local in space, corresponding to a large effective mass at time $t=0^+$, independent of the quench protocol.  

\section{Long time limit of the Green's functions}
\label{app:long-time-limit}
We can extend the analyses for the equal time Keldysh function to different time arguments $t,t'$, but still in the limit
\begin{equation}
 q^zt/\gamma^{z/2}\gtrsim q^zt'/\gamma^{z/2}\gg 1 \text{ and }q^z(t-t')/\gamma^{z/2}\ll1. 
\end{equation}
We use the long time expansion for $\phi\simeq \phi_{\text{eq}}-\int_{t_\gamma}^t dsG^R_{\text{eq}}(t-s) r(s)\phi_{\text{eq}}(s)$ to obtain:
\begin{align}
 G_r^K(q,t,t')=& -i\langle[\phi(t),\phi(t')]_+ \rangle\nonumber\\
  =&G^K_{\text{eq}}(q,t-t')- \int_{t_\gamma}^t ds G^R_{\text{eq}}(t-s) r(s) G_{\text{eq}}^K(s,t')\nonumber\\
 &- \int_{t_\gamma}^{t'}ds G_{\text{eq}}^R(t'-s) r(s) G_{\text{eq}}^K(s-t)\label{app:eq:GK}
 \end{align}
To evaluate the $s$-integration we express $G^R$ by its Laplace-transformation and $G^K$ by its Fourier transformation:
\begin{align}
 & - \int_{}^t ds G^R_{\text{eq}}(t-s) r(s) G_{\text{eq}}^K(s,t')\nonumber\\
 =& - i \int \frac{d\omega d\omega'}{2\pi^2}{\rm Im} G_{\text{eq}}^R(\omega) G_{\text{eq}}^K(\omega')\int^t ds r(s) e^{-i\omega (t-s)-i\omega'(s-t')}
 \end{align}
 Here, $\omega$ and $\omega'$ are of order of the typical mode-frequency $q^{-z}\gamma^{z/2}$. In the limit where the time difference $t-t'$ is small, but $t, t'$ are both large compared to the mode-frequency the $s$-integral can be approximated by
\begin{equation}
 \int_{t_\gamma}^tdsr(s)e^{-i\omega(t-s)-i\omega'(s-t')}\simeq r\left(t\right)\frac{e^{-i\omega'(t-t')}}{i(\omega-\omega')}
\end{equation}
In a next step, we use Kramers-Kronig relation to preform one frequency-integration. This finally yields
 \begin{align}
 & - \int_{}^t ds G^R_{\text{eq}}(t-s) r(s) G_{\text{eq}}^K(s,t')\nonumber\\
 =&- r(t)\int \frac{d\omega'}{2\pi}{\rm Re} G_{\text{eq}}^R(\omega') G_{\text{eq}}^K(\omega') e^{-i\omega'(t-t')}
\end{align}
The same procedure leads for the second integral in Eq.~\eqref{app:eq:GK} :
\begin{align}
 &- \int^{t'}ds G_{\text{eq}}^R(t'-s) r(s) G_{\text{eq}}^K(s-t)\nonumber\\
 &\simeq -r(t')\int \frac{d\omega'}{2\pi}{\rm Re} G_{\text{eq}}^R(\omega') G_{\text{eq}}^K(\omega') e^{-i\omega'(t-t')}
\end{align}
For $r(t)\approx r(t')\approx r(t_a)$ with $t_a=(t+t')/2$ this yields to the result given in Eq.~\eqref{eq:long-time-GK-tt'} in the main text.
In analogy to the long time limit of $G_r^K(q,t,t)$ we can derive a similar behavior for the retarded Greens function $G_r^R(q,t,t')$.
It follows with the long time expansion for $\phi$ and the definition of $G_r^R$ in Eq.~\eqref{eq:defGR} that
\begin{align}
 &G_r^R(t,t')=-i\theta(t-t')\langle[\varphi(t),\varphi(t')]_- \rangle\nonumber\\
 &\simeq  G^R_{\text{eq}}(t-t')\nonumber\\
 &+i\theta(t-t')\left(\int_0^tds G^R_{\text{eq}}(t-s) r(s) \langle[\varphi(s),\varphi(t')]_-\rangle\right.\nonumber\\
 &\left.+ \int_0^{t'}ds \langle[\varphi(t),\varphi(s)]_- \rangle r(s)  G^R_{\text{eq}}(t'-s)\right)\nonumber\\
 =&G^R_{\text{eq}}(t-t')-\int_{t'}^tds G^R_{\text{eq}}(t-s) r(s) G^R_{\text{eq}}(s-t').
\end{align}
We express the equilibrium $G_r^R$ by their Laplace-transformation and obtain
\begin{align}
 \delta G_r^R(t,t')=&-\int_{t'}^tds G^R_{\text{eq}}(t-s) r(s) G^R_{\text{eq}}(s-t')\nonumber\\
 =&\int \frac{d \omega d\omega'}{\pi^2}{\rm Im} G_{\text{eq}}^R(\omega){\rm Im} G_{\text{eq}}^R(\omega')\nonumber\\
 &\times\int_{t'}^tds\frac{a\gamma}{s^{2/z}} e^{-i\omega(t-s)-i\omega'(s-t')}
\end{align}
$\omega$ and $\omega'$ are of order of the typical mode-frequency $q^{-z}\gamma^{z/2}$. In the limit where the time difference $t-t'$ is small, but $t, t'$ are both large compared to the mode-frequency the $s$-integral can be approximated by
\begin{align}
 &\int_{t'}^tdsr(s)e^{-i\omega(t-s)-i\omega'(s-t')}\nonumber\\
 &\simeq r\left(\frac{t+t'}{2}\right)\frac{1}{i(\omega-\omega')}\left(e^{-i\omega'(t-t')}-e^{-i\omega(t-t')} \right)
\end{align}
This yields to the result given in Eq.~\eqref{eq:long-time-GR-tt'}
\begin{align}
 &\delta G_r^R(t,t')\simeq -i r\left(\frac{t+t'}{2}\right) \int \frac{d \omega d\omega'}{\pi^2}\frac{{\rm Im} G_{\text{eq}}^R(\omega){\rm Im} G_{\text{eq}}^R(\omega')}{\omega-\omega'}\nonumber\\
 &\times\left(e^{-i\omega'(t-t')}-e^{-i\omega(t-t')}\right)\nonumber\\
 &= -2i r\left(\frac{t+t'}{2}\right)    \int \frac{d\omega'}{\pi} {\rm Re} G_{\text{eq}}^R(\omega'){\rm Im} G_{\text{eq}}^R(\omega')e^{-i\omega'(t-t')}\nonumber\\
 &=- 4i r\left(\frac{t+t'}{2}\right) C^R(t-t').
\end{align}


%

\end{document}